\documentclass[twocolumn,prb,superscriptaddress,floatfix]{revtex4-1}
\usepackage[dvipsnames]{xcolor}
\usepackage{feynmf}
\usepackage{tikz-feynman}
\tikzfeynmanset{compat=1.0.0}
\tikzfeynmanset{/tikzfeynman/momentum/arrow shorten = 0.3}
\tikzfeynmanset{/tikzfeynman/warn luatex = false}
\usepackage{amssymb}
\usepackage{amsbsy}
\usepackage{amsmath}
\usepackage{bm}
\usepackage{textcomp}
\usepackage{placeins}
\usepackage{gensymb}
\usepackage[breaklinks,colorlinks = true,linkcolor = blue,urlcolor=cyan,citecolor=blue]{hyperref}
\usepackage{multirow}
\usepackage{array}
\usepackage{booktabs}
\usepackage{upgreek}
\usepackage{tabularx}
\usepackage{epsfig,psfrag,amsopn}
\usepackage{mathrsfs}
\usepackage{float}
\usepackage{graphicx}
\usepackage{subcaption}

\begin{document}

\title{Probing signatures of fractionalisation in candidate quantum spin liquid \texorpdfstring{Cu$_2$IrO$_3$}~ via anomalous Raman scattering}

\author{Srishti Pal}
\affiliation{Department of Physics, Indian Institute of Science, Bengaluru 560012, India}

\author{Arnab Seth}
\affiliation{International Centre for Theoretical Sciences, Tata Institute of Fundamental Research, Bengaluru 560089, India}

\author{Piyush Sakrikar}
\affiliation{Indian Institute of Science Education and Research (IISER) Mohali, Knowledge City, Sector 81, Mohali 140306, India}

\author{Anzar Ali}
\affiliation{Indian Institute of Science Education and Research (IISER) Mohali, Knowledge City, Sector 81, Mohali 140306, India}

\author{Subhro Bhattacharjee}
\affiliation{International Centre for Theoretical Sciences, Tata Institute of Fundamental Research, Bengaluru 560089, India}

\author{D. V. S. Muthu}
\affiliation{Department of Physics, Indian Institute of Science, Bengaluru 560012, India}

\author{Yogesh Singh}
\affiliation{Indian Institute of Science Education and Research (IISER) Mohali, Knowledge City, Sector 81, Mohali 140306, India}

\author{A. K. Sood}
\email[E-mail:~]{asood@iisc.ac.in}
\affiliation{Department of Physics, Indian Institute of Science, Bengaluru 560012, India}

\date{\today}

\begin{abstract}

{Long-range entanglement in quantum spin liquids (QSLs) lead to novel low energy excitations with fractionalised quantum numbers and (in 2D) statistics.  Experimental detection and manipulation of these excitations present a challenge particularly in view of diverse candidate magnets. A promising probe of fractionalisation is their coupling to phonons. Here we present Raman scattering results for the $S = 1/2$ honeycomb iridate Cu$_2$IrO$_3$, a candidate Kitaev QSL with fractionalised Majorana fermions and Ising flux excitations. We observe anomalous low temperature frequency shift and linewidth broadening of the Raman intensities in addition to a broad magnetic continuum both of which, we derive, are naturally attributed to the phonon decaying into itinerant Majoranas. The dynamic Raman susceptibility marks a crossover from the QSL to a thermal paramagnet at $\sim$120 K. The phonon anomalies below this temperature demonstrate a strong phonon-Majorana coupling. These results provide  for evidence of spin fractionalisation in Cu$_2$IrO$_3$.}

\end{abstract}

\maketitle

\section{Introduction}

{Recent advances in condensed matter physics and material sciences have shown that several so-called {\it elementary particles}, originally conceived in context of high energy physics can emerge as low energy excitations (quasi-particles) in condensed matter systems. In addition to providing an impetus to the paradigm of emergent quantum phenomena~\cite{Anderson,laughlin2000cover,RevModPhys.89.041004}, these materials then provide concrete contexts to understand the properties of these novel excitations and the settings for their emergence as an interplay of symmetries and many-body entanglement. This ranges from the weakly correlated physics of Dirac fermions in monolayer graphene \cite{castro2009electronic,novoselov2005two} and Weyl fermions in topological semimetals~\cite{PhysRevB.83.205101,Yan,RevModPhys.90.015001} on one hand to the strongly correlated fractionalised excitations in fractional quantum Hall systems~\cite{read2000paired,nayak2008non}.}

{In this context, the possibility of emergence of the elusive (in high energy particle physics) Majorana fermion \cite{Wilczek,alicea2012new} in several candidate solid state systems like topological superconductors \cite{kitaev2001unpaired,Qi,PhysRevB.73.220502,das2012zero,mourik2012signatures}, fractional quantum Hall systems~\cite{banerjee2018observation}, and QSLs~\cite{Kitaev,kasahara2018majorana,Banerjee1,PhysRevB.85.094418} have been invoked to account for startling novel low energy properties of these systems. Among them, the Kitaev QSL \cite{Kitaev} on the isotropic honeycomb lattice provides a unique opportunity where propagating Majorana excitations coupled to emergent $Z_2$ fluxes arise due to the long-range quantum entanglement present in the system resulting in the fractionalisation of the underlying microscopic spin-$\frac{1}{2}$s~\cite{Kitaev}.}

{Our present understanding suggests that a key ingredient in realising Kitaev physics is specific {\it compass} spin-spin interactions~\cite{Jackeli,RevModPhys.87.1} on a tri-coordinated motif consisting of edge-sharing octahedra~\cite{Jackeli,PhysRevLett.112.077204}. Our growing understanding of magnets with strong spin-orbit coupling have provided a slew of such candidate Kitaev QSL materials containing 4d and 5d transition metal ions. The most notable ones among these are Na$_2$IrO$_3$~\cite{Mehlawat} and $\alpha$-RuCl$_3$~\cite{Do} on a two-dimensional layered honeycomb motif and $\gamma$- and $\beta$-Li$_2$IrO$_3$~\cite{Glamazda} on a three-dimensional hyper-honeycomb lattice. A combination of thermodynamic measurements and scattering experiments~\cite{Banerjee1,Banerjee2} on these ``first-generation" Kitaev materials show extremely interesting finite temperature behaviour including possible signatures of fractionalisation and thereby raise questions about their proximity to a Kitaev (or other) QSLs~\cite{Banerjee1,Winter,kasahara2018majorana}. However, they ultimately order magnetically at a much lower temperature possibly due to additional non-Kitaev interactions in these systems. Thus, while the above compounds are very interesting in their own rights to understand the possible interplay of magnetic fluctuations and fractionalisation, the realisation of the Kitaev QSL with pristine signatures of the fractionalised Majoranas still remains an open issue.}

{In this paper, we report our results on the Raman scattering and magneto-elastic coupling of the ``second-generation" Kitaev material, Cu$_2$IrO$_3$~\cite{Abramchuk}, where the magnetic order is absent suggesting the possibility of smaller non-Kitaev interactions. In particular, despite a Curie-Weiss temperature and effective magnetic moment similar to Na$_2$IrO$_3$, $\mu$SR and specific heat studies on Cu$_2$IrO$_3$ have shown an absence of magnetic order and an excitation spectrum dominated by low-energy Ir spin dynamics~\cite{Choi, Kenney}. The correlated nature of this low temperature dynamic paramagnet is further supported by the NQR measurements~\cite{Takahashi}. These findings suggest that Cu$_2$IrO$_3$ may offer an ideal playground to investigate fractionalisation in a Kitaev QSL with an eye towards positive characterisation of the Majorana fermions therein. Such Majorana fermions can then couple to the optical phonons through the regular spin-phonon coupling leading to their characteristic experimental signatures. Incidentally such coupling has only been studied for the acoustic phonons for a Kitaev QSL~\cite{Ye,Serbyn,Metavitsiadis}.}

{Indeed, strong spin-orbit coupling results in intricate mixing of the real and spin space. Thus, on generic grounds one expects these compounds to have enhanced spin-lattice coupling. Can this spin-(optical) phonon coupling, as probed in Raman scattering, then lead to positive identification of the possible fractionalised excitations (Majorana fermions in a Kitaev QSL) in candidate QSL materials such as Cu$_2$IrO$_3$? The central result of our work is the anomalous shift and broadening of the Raman active phonons in Cu$_2$IrO$_3$ (Fig.~\ref{fig2}) which indicate extra decay channels become active at low temperatures for the low energy Raman active phonons. In absence of magnetic order, a natural candidate for coherent modes that can result in phonon decay, within Kitaev phenomenology, are itinerant Majorana fermions. Indeed, we find that such fractionalisation does provide a successful explanation for our Raman measurements. Our estimate of the Kitaev exchange from the band edge of the magnetic continuum is consistent with earlier NQR measurements. The temperature dependence of Raman susceptibility is non-monotonic and clearly evidences fermionic Majorana excitations prevailing over a conventional Bosonic background below about $120$~K.}

{The ``second-generation" Kitaev materials, such as Cu$_2$IrO$_3$, are obtained by partially or fully replacing the alkali atoms in $\alpha$-$A_2$IrO$_3$ with other atoms. Incredibly, the new materials produced in this manner, H$_3$LiIr$_2$O$_6$~\cite{Kitagawa}, Cu$_2$IrO$_3$~\cite{Abramchuk}, and Ag$_3$LiIr$_2$O$_6$~\cite{Bahrami}, have been shown to be QSL candidates with no signatures of magnetic order using various thermodynamic and dynamic probes~\cite{Kitagawa, Choi, Bahrami, Takahashi}. In these second generation Kitaev materials, the edge-sharing IrO$_6$ octahedra forming a honeycomb lattice plane is retained. However, the interplanar connectivity is changed. For example, Cu$_2$IrO$_3$ crystallises in the same \textit{C2/c} monoclinic structure as Na$_2$IrO$_3$. The honeycomb layers are formed by an edge-sharing (Ir$_{2/3}$Cu$_{1/3}$)O$_6$ octahedral arrangement identical to Na$_2$IrO$_3$, but the interlayer connections via distorted NaO$_6$ octahedra are replaced by linear CuO$_2$ dumbbells resulting in a larger $c$-axis. This enhanced 2D character of the honeycomb layers along with the proximity of the Ir-Ir-Ir angles towards the ideal value of 120$\degree$ compared to its predecessor Na$_2$IrO$_3$, puts copper iridate closer to the ideal Kitaev limit. Similar interlayer bonding is found for H$_3$LiIr$_2$O$_6$~\cite{Kitagawa} and Ag$_3$LiIr$_2$O$_6$~\cite{Bahrami}.}

{The nature of the synthesis makes these second generation Kitaev materials prone to disorder.  For example, proton positional disorder in H$_3$LiIr$_2$O$_6$, Ag positional disorder in Ag$_3$LiIr$_2$O$_6$, or Cu mixed valent disorder in Cu$_2$IrO$_3$. The possible role of disorder in stabilising the QSL in these materials has been discussed recently\cite{Choi, Kitagawa, Yadav, Li, JKnolle}. In the context of Raman measurements, the proton disorder in H$_3$LiIr$_2$O$_6$ has been suggested to lead to the observed anomalously broad phonon modes as well as the weak fermionic contribution to the magnetic continuum~\cite{Pei}. Unlike H$_3$LiIr$_2$O$_6$, however, the synthesis of Ag$_3$LiIr$_2$O$_6$ and Cu$_2$IrO$_3$ can be controlled to reduce the disorder.  For example, while disordered Ag$_3$LiIr$_2$O$_6$ samples with Ag randomly occupying voids of the LiIr$_2$O$_6$ honeycomb layers show broad features in the heat capacity at low temperatures, higher quality samples which do not have this Ag positional disorder actually undergo long-ranged magnetic order~\cite{Bahrami2}. For Cu$_2$IrO$_3$ as well, disorder can lead to measurable consequences. Initial reports observed a glassy state at low temperatures (\textit{T} $\sim$3.5~K) using magnetic susceptibility~\cite{Abramchuk}. A careful experimental and theoretical study including x-ray absorption spectroscopy, $\mu$SR, and density functional theory identified the source of disorder as an $8.5$ to $13\%$ contamination of Cu$^{2+}$ instead of Cu$^{+}$ in Cu$_2$IrO$_3$~\cite{Kenney}. The Cu$^{2+}$ spins were found to be located within the voids of the honeycomb layers formed by the edge-sharing IrO$_6$ octahedra. From $\mu$SR, the Cu$^{2+}$ spins were found responsible for the spin-glass freezing while the Ir$^{4+}$ honeycomb sublattice was found to be in a dynamically fluctuating QSL-like state.  Additionally, the frozen regions and the QSL regions were found to occupy different volume fractions of the sample. Thus, it is clear that a majority ($\sim 90\%$) of the disordered Cu$_2$IrO$_3$ sample is actually in a QSL state and the frozen volume fraction is phase separated from the QSL part of the material \cite{Kenney}.}

{Subsequently, we have been able to synthesise high quality Cu$_2$IrO$_3$ samples-- used for the present experiments-- which do not show any features of a glassy state in ac magnetic susceptibility (see Fig.~\ref{fig_a0}(d)). This strongly suggests that our Cu$_2$IrO$_3$ samples have a much smaller amount of disorder. This is confirmed by microscopic probes of magnetism, like $\mu$SR which show an absence of any static magnetism down to $260$~mK in the present samples~\cite{Choi}. These measurements additionally show that the Ir spins stay in a dynamically fluctuating QSL-like state down to temperatures more than two orders of magnitude smaller than the Curie-Weiss temperature~\cite{Choi}. Even in our higher quality samples, however, some small amount of disorder remains, as evidenced by the low temperature ($T \leq 20$~K) sub-Curie law susceptibility (see Fig.~\ref{fig_a0}(c) and (e)) and scaling behaviours in various thermodynamic quantities which is consistent with a small fraction of random-singlets in a background of a QSL like phase~\cite{Choi,sanyal2020emergent}. The absence of freezing in our samples suggests a smaller concentration of Cu$^{2+}$ impurities. Analysis of the low temperature susceptibility demonstrates that $\approx 5\%$ of $S = 1/2$ impurities can explain the low temperature sub-Curie law (see Appendix~\ref{A_A}).}

{We have therefore chosen the second generation Kitaev material Cu$_2$IrO$_3$ to study the fractionalisation predicted for a Kitaev QSL-- most prominently the itinerant Majorana fermions~\cite{Motome}. Neutron scattering on iridates is difficult because of the strong absorption of neutrons by iridium, although some efforts to measure iridates using special experimental setups have been reported~\cite{SKChoi}. An important and complementary experimental route to probe the novel fractionalised excitations is provided by Raman spectroscopy. Importantly, Raman signatures contain, as we show below, two different but related signatures of the low energy Majorana fermions-- (1) via their direct coupling to photons leading to broad magnetic continuum~\cite{Perreault,Nasu2,Sandilands}, and (2) the additional decay of Raman active phonons through their coupling to Majorana excitations via spin-phonon coupling. Positive identification of Majorana signatures in both of these aspects, we show, strongly suggests the relevance of Kitaev QSL physics in Cu$_2$IrO$_3$ with low energy Majorana fermions. Indeed, in Raman scattering for $\gamma$- and $\beta$-Li$_2$IrO$_3$~\cite{Glamazda}, $\alpha$-RuCl$_3$~\cite{Sandilands, Wulferding, Glamazda2} and H$_3$LiIr$_2$O$_6$~\cite{Pei}, a broad magnetic continuum has been detected in the low-energy Raman profile. Even though $\gamma$- and $\beta$-Li$_2$IrO$_3$ and $\alpha$-RuCl$_3$ have magnetically ordered ground states, the temperature evolution of the magnetic background is typified by dominant Fermi statistics and has been attributed to the fractionalised Majorana fermions~\cite{Glamazda,Sandilands,Glamazda2,Wulferding}.}

\section{Experimental Methods}

{High quality polycrystalline samples of Cu$_2$IrO$_3$ were prepared by a low temperature topotactic reaction of Na$_2$IrO$_3$ with CuCl as reported previously~\cite{Choi}. Powder X-ray diffraction confirmed the expected crystal structure (C2/c space group) and ac and dc susceptibility measurements down to $300$~mK (Appendix~\ref{A_A}) confirmed the absence of spin-freezing which has been reported to contaminate the low temperature magnetism for some Cu$_2$IrO$_3$ materials reported previously~\cite{Abramchuk,Takahashi}.}

{The unpolarised Raman spectra at room temperature were recorded in a backscattering geometry using Horriba LabRAM HR Evolution Spectrometer equipped with a thermoelectric cooled charge coupled device (CCD) (HORIBA Jobin Yvon, SYNCERITY 1024 X 256). The low temperature Raman measurements were performed from 6 K to 295 K with 532 nm DPSS laser illuminating the sample with less than $\sim$1.5 mW power. Temperature variation was done with closed cycle He cryostat (Cryostation S50, Montana Instruments) with a temperature stability of $\approx \pm$1 K. The cross-polarised Raman spectra were recorded using Horriba LabRAM HR Evolution Spectrometer with a ULF (ultra low-frequency) set up to record spectrum down to 10 cm$^{-1}$. The low temperature Raman measurements on that set up were performed from 4 K to 300 K using continuous flow liquid helium cryostat (MicrostatHe2 by Oxford Instruments) with a temperature stability of $\approx \pm$1 K.}

\section{Experimental Results}

\begin{figure}%[h!]
\centering
\includegraphics[width=8.6cm]{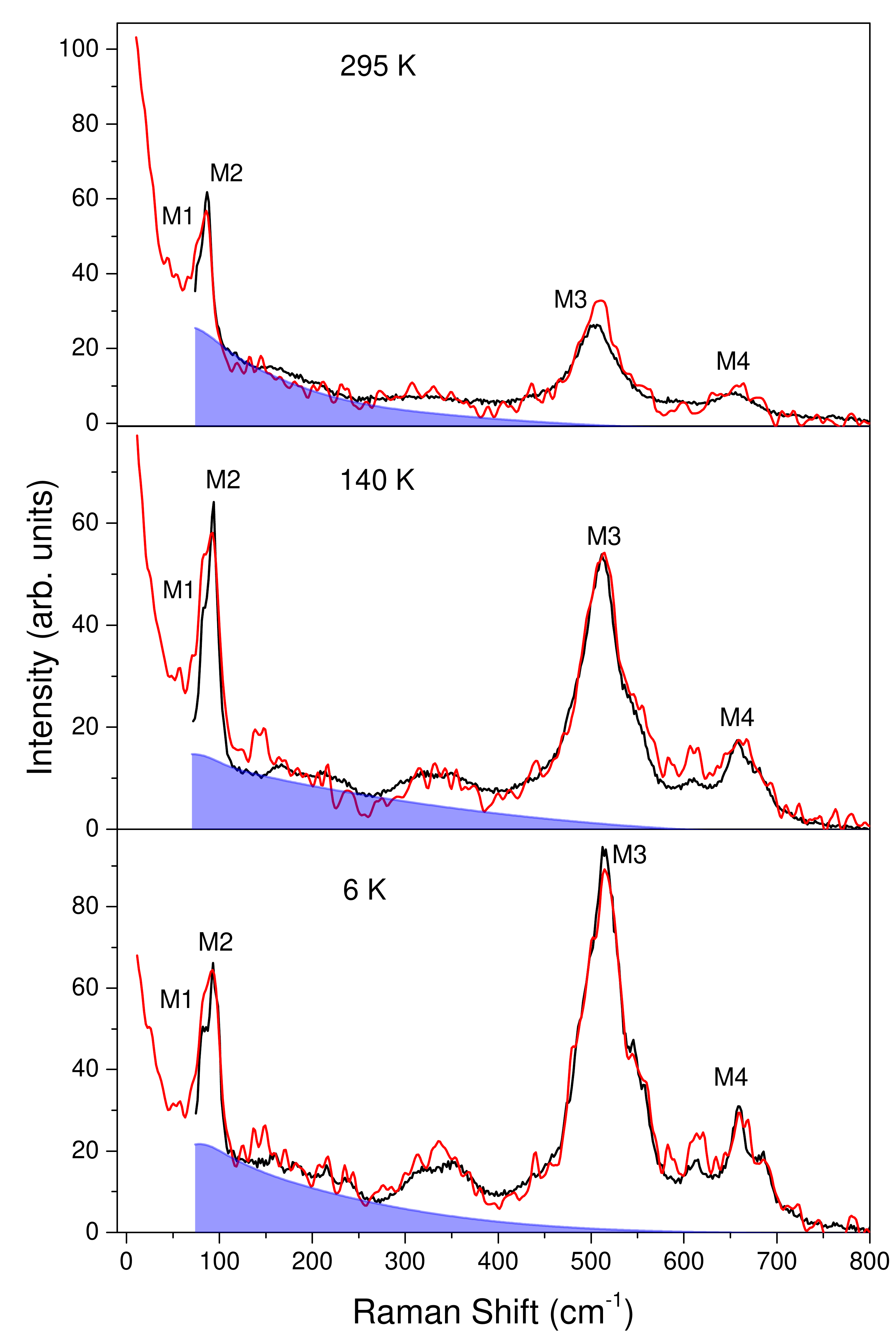}
\caption{\small Black and red lines represent the unpolarised and cross-polarised Raman spectra of Cu$_2$IrO$_3$ at three different temperatures in the spectral ranges of 70 - 800 cm$^{-1}$ and 10 - 800 cm$^{-1}$, respectively. The shaded region represents the low-energy magnetic continuum.}
\label{fig1}
\end{figure}

{Fig.~\ref{fig1} shows unpolarised and cross-polarised (to avoid any contribution from Rayleigh scattering at low frequencies) Raman spectra of Cu$_2$IrO$_3$ at a few representative temperatures with sharp phonon modes and a quasi-elastic scattering (QES) component (linewidth $\sim$50 cm$^{-1}$) superimposed on a broad continuum extending up to $\sim$600 cm$^{-1}$. As observed experimentally in other Kitaev materials $\alpha$-RuCl$_3$~\cite{Sandilands, Glamazda2, Wulferding} and $\gamma$- and $\beta$-Li$_2$IrO$_3$~\cite{Glamazda}, phonons are superimposed on a broad background which is temperature dependent. This broad Raman background in experiments has been attributed to the gapless itinerant Majorana fermions of a Kitaev QSL. Finite temperature simulations for the pure Kitaev model by Nasu et al.~\cite{Nasu2} reproduce the broad continuum with a band edge extending up to $\simeq3\mid$\textit{J}$_K\mid$ (arising from the Majorana fermion bandwidth) where \textit{J$_K$} is the Kitaev coupling strength. As shown in Fig.~\ref{fig1}, the upper cut-off of the magnetic continuum in Cu$_2$IrO$_3$ gives an experimental estimate for $\mid$\textit{J}$_K\mid$ $\approx$ 24 meV, in good agreement with recent estimates (17 to 30 meV) from the low-energy spin excitation gap seen in NQR studies~\cite{Takahashi}. This intriguing broad magnetic continuum (Fig.~\ref{fig1}) then begs for a careful closer investigation - a topic to which we shall now focus on. This will be followed by the discussion of the low energy quasi-elastic signal, whose weight, appears to generically diminish at low temperature (see below).}

\begin{figure}%[h!]
\centering
\includegraphics[width=8.6cm]{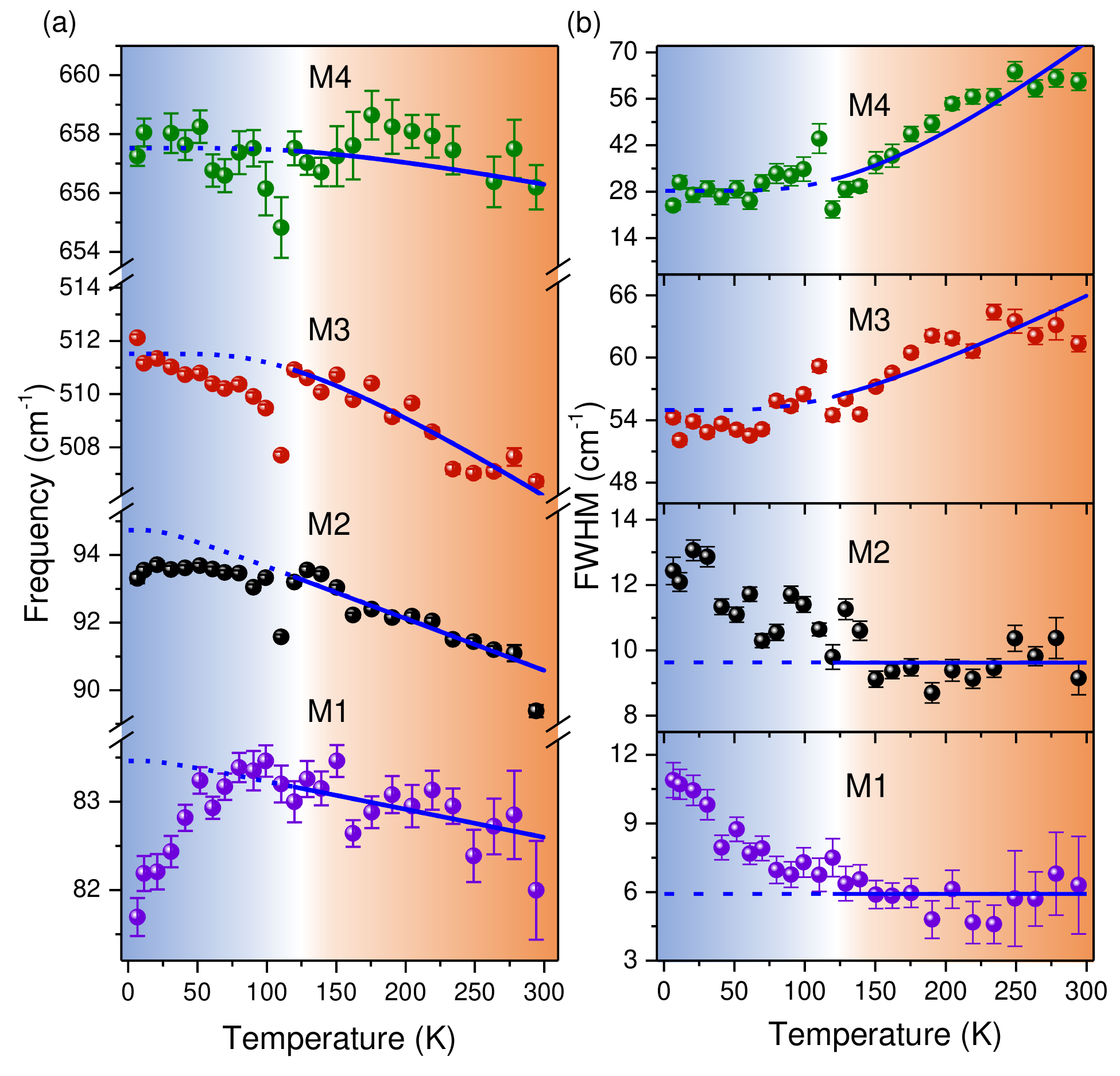}
\caption{\small  Temperature dependence of ({\bf a}) phonon frequency and ({\bf b}) FWHM. Blue lines are anharmonic fits to phonon frequencies and FWHMs. Shaded regions demonstrate the boundary between the normal and the Kitaev paramagnetic states.}
\label{fig2}   
\end{figure}

\subsection{Anomalies in frequencies and linewidths of Raman active phonons}

{To further probe the signatures consistent with fractionalisation of the spins into Majorana fermions, we now look for the effect of Majorana excitations on phonons, if any, especially at low temperatures $(T\lesssim J_K$). Such effect should be particularly strong for the phonons embedded/close to the magnetic continuum.}

{Of the 39 active Raman modes expected for monoclinic (\textit{C2/c}) Cu$_2$IrO$_3$ ($\Gamma_{Raman}$ = 18$A_g$ + 21$B_g$), 13 modes could be detected at $\sim$6 K in the frequency range 70-800 cm$^{-1}$ (8.7 - 99.2 meV).  All the phonon modes are fitted with symmetric Lorentzian profile function for the entire range of temperature. The overall phonon spectrum remains almost unchanged with increasing temperature except for the thermal broadening of weaker modes making them undetectable at higher temperatures. No change in the number of Raman modes confirm the stability of the ambient crystal symmetry down to 6 K. This is an advantage that Cu$_2$IrO$_3$ has over $\alpha$-RuCl$_3$, which undergoes a structural transition around $140$~K, further obscuring attempts to establish connections between the onset of Majorana fermions and phonons~\cite{Glamazda2}.}

{Normally, a monotonic temperature dependence of phonon parameters is expected because the phonon self-energies are typically determined by lattice anharmonicity which reduces monotonically with temperature~\cite{Menendez}. This is however not the case in Cu$_2$IrO$_3$ with anomalous temperature evolution of frequencies and FWHMs of the phonon modes below $T\simeq 120$ K. The temperature dependence of the frequency and FWHMs of the strong phonon modes (marked M1, M2, M3 and M4 in Fig.~\ref{fig1} and Fig.~\ref{fig_a3}) are shown in Fig.~\ref{fig2}(a)-(b). The solid blue lines are the fit from 295 K to 120 K to the simple cubic anharmonic model~\cite{Klemens} representing phonon (frequency $\omega$) decaying in two phonons of equal frequency ($\omega$/2) (see Appendix~\ref{A_E} for fitting details). The dashed lines are extrapolations of the fits to lower temperatures. The frequencies (FWHM) are lower (higher) than expectation from the cubic anharmonic temperature dependence of phonons. The latter in particular is suggestive of extra channels provided by the magnetic continuum for the Raman active phonons to decay. This effect though most dramatic for M1 is seen for all modes below $T\lesssim 120$ K $\approx 0.43~J_K$. The above anomaly is very much different from the phonon softening in magnetically ordered materials, such as Fe$_3$GeTe$_2$~\cite{du2019lattic}, where similar anomalies are associated with the magnetic order. For Cu$_2$IrO$_3$, however, no such magnetic order is present down to the lowest temperature measured. At this point we note that, an estimated small fraction ($\approx 5\%$) of random-singlets emerging below $\sim$20 K (see Appendix~\ref{A_A}) is incongruous to induce any anomaly in the phonon modes at the much higher temperature scale of 120 K.}

{Remarkably, the numerical studies~\cite{Nasu, Nasu2} of the pure Kitaev model found that such a temperature scale, $T_h\sim 0.4-0.5~J_K$, is associated with the completion of transfer of spectral weight of a coherent itinerant Majorana fermion to an incoherent one. Indeed, the above temperature is associated with the van-Hove singularity of the free Majorana dispersion in the zero-flux sector whose depletion is completed at \textit{T} $\sim$\textit{T$_h$}. The above agreement of $T_h$ of the pure Kitaev model numerics is seen for all the Raman-active modes. At this point, we note that for the Kitaev QSL there is another energy-scale $T_l\sim $0.012-0.015\textit{J$_K$} associated with the $Z_2$ fluxes~\cite{Kitaev,Nasu,Nasu2}. Although such low temperature is not accessible to the current experiment, the  frequency (FWHM) of the phonon decreases(increases) monotonically till the lowest accessible temperature, \textit{T} $\sim$6 K.}

\subsection{Intensity of $M3$ mode}

{In absence of spin-lattice coupling, the temperature dependence of integrated intensities of Raman phonons should follow the conventional Bose-Einstein  distribution. The high-frequency $M3$ mode ($\hbar\omega_{M3} \approx$ 63 meV) shows a strong departure from the above expectation and shows a strong enhancement of intensity with decreasing temperatures as seen in Fig.~\ref{fig1}. In fact, we find (see Appendix~\ref{A_F}) that the intensity of $M3$ closely follows the temperature dependence of the DC susceptibility and thus, is dependent on the spin-spin correlation. The susceptibility, in turn, shows clear deviation from the high-temperature Curie-Weiss (CW) behaviour below $\sim$120 K. Such anomalous behaviour can arise from transfer of magnetic dipole spectral weight to the phonons via spin-lattice coupling~\cite{Allen}. Indeed, the phonon intensities are expected to depend on the spin-spin correlations \cite{suzuki1973theory}. This reiterates the presence of sizeable spin-lattice coupling in Cu$_2$IrO$_3$.}

\subsection{Low-energy magnetic continuum}

{Within the Kitaev QSL phenomenology, which presently forms the natural framework to understand the anomalous Raman scattering, we attribute the low energy magnetic continuum to that of the Majorana fermions scattering from the $Z_2$ fluxes. In this regard, as shown in Ref.~\onlinecite{Nasu2}, the primary contribution to the magnetic continuum arises from the itinerant Majorana fermions interacting with the Raman photon while the effect of the low energy $Z_2$ fluxes on the Majorana fermions inside the QSL is to renormalise the fermion bandwidth and density of states~\cite{Nasu,Nasu2}. As discussed in Ref.~\onlinecite{Nasu2}, there are two distinct contributions dominating over distinct energy scales. While at high energies the two-fermion creation processes dominate, at low energies simultaneous creation and annihilation of two fermions contribute substantial weight to the (integrated) Raman intensity.}

\begin{figure}[h!]
\centering
\includegraphics[width=8.6cm]{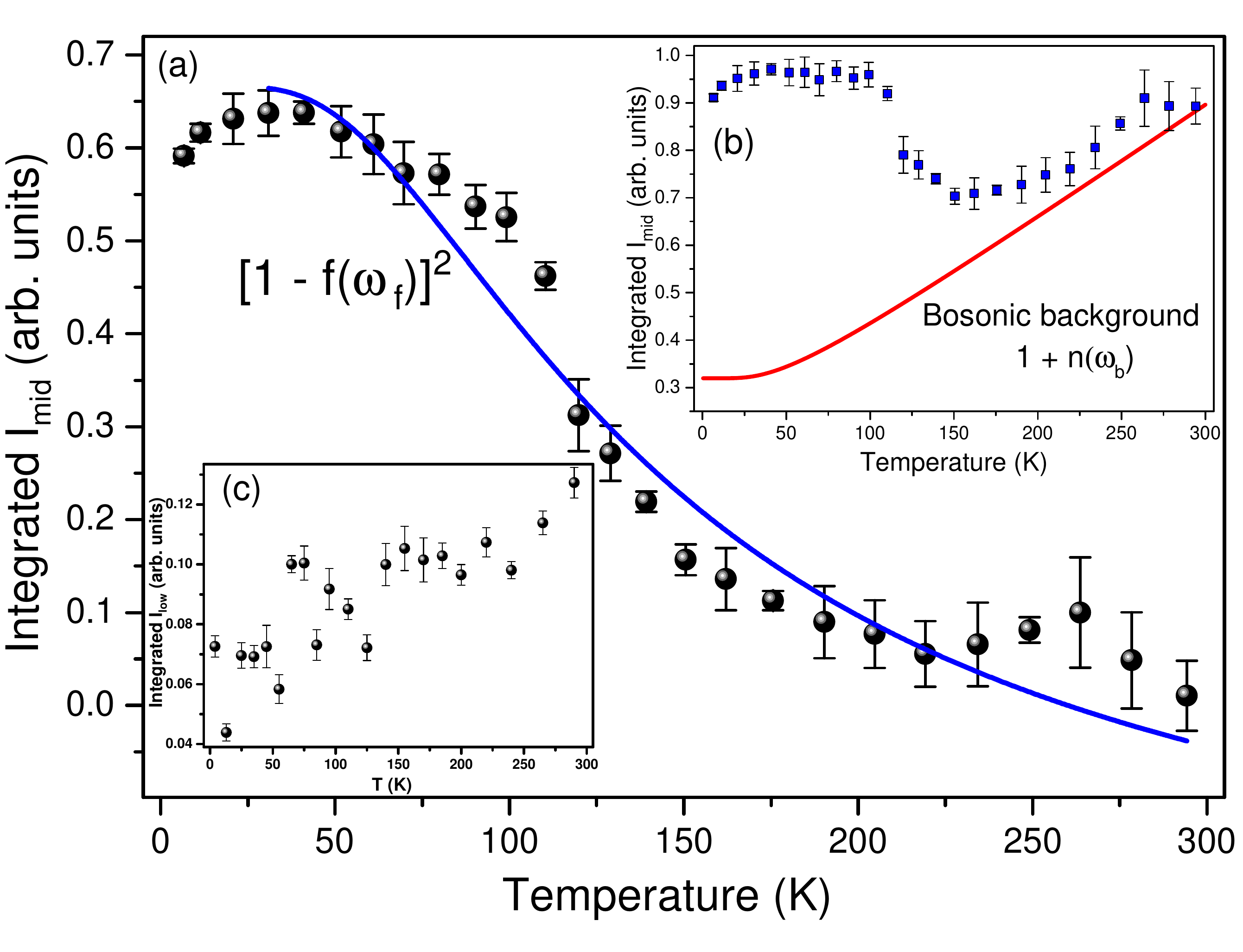}
\caption{\small ({\bf a}) Symbols denote the magnetic contribution to integrated \textit{I$_{mid}$} in the frequency range 120-260 cm$^{-1}$ after subtracting the bosonic background (shown in ({\bf b})). The blue solid curve represents fitting by the two-fermion scattering form $A + B(1-f)^2$, with $f=1/(1 + e^{\hbar\omega/k_B T})$ being the Fermi distribution function. ({\bf c}) Symbols denote the magnetic continuum integrated in the low-frequency region of 10-45 cm$^{-1}$.}
\label{fig3}
\end{figure}

{We particularly focus on the former mid to high energy contribution, the details of which (see below) are relatively more robust than the low energy signatures as discussed below. To this end, following Ref.~\onlinecite{Nasu2}, in order to extract Majorana fermion energy scale from the low-energy continuum, following~\cite{Nasu2}, the Raman intensity \textit{I($\omega$)} is integrated over the intermediate frequency range of $\omega_{min} < \omega < \omega_{max}$ to obtain $I_{mid} = \int_{\omega_{min}}^{\omega_{max}} I(\omega)d\omega$. The temperature dependence of \textit{I$_{mid}$} in the frequency interval 120-260 cm$^{-1}$ is plotted in Fig.~\ref{fig3}(b). As is clear from the inset (b), \textit{I$_{mid}$} has a non-monotonic temperature dependence with the high temperature regime dominated by the standard one-particle scattering due to thermal Bose factor $[1 + n(\omega_b)] = 1/(1 - e^{-\hbar\omega_b/k_B T}$), with $\omega_b$ = 11 meV, extracted from the Bosonic fit as a fitting parameter. This Bosonic background is attributed to phonons since the system does not entertain other Bosonic excitations like magnons due to lack of long-range spin ordering down to lowest measurable temperature. A confirmation of this is obtained from the fact that the value of $\omega_b$ matches well with the strongest phonon mode at $\sim$92 cm$^{-1}$ in the low-frequency regime.}

{Fig.~\ref{fig3}(a) (the main panel) demonstrates the temperature evolution of integrated \textit{I$_{mid}$} after subtracting the non-magnetic Bosonic background. The magnetic contribution to \textit{I$_{mid}$} enhances significantly below 120 K as clearly indicated by deviation from thermal behaviour and can be well fitted to the form~\cite{Nasu2} $A + B[1 - f(\omega_f)]^2$ with $\omega_f$ = 19 meV, where $f(\omega_f) = 1/(1 + e^{\hbar\omega_f/k_B T}$) is the Fermi distribution function. This typical scaling of \textit{I$_{mid}$}, as mentioned above, is associated with the scattering contribution from the process of creation or annihilation of Majorana fermion pairs~\cite{Nasu2}. The Majorana energy scale for Cu$_2$IrO$_3$ is deduced from the fermionic fit with $\omega_f$ = 19 meV ($\approx$ 0.8$\mid$\textit{J}$_K\mid$) and is in accordance with a Kitaev QSL phase considering similar energy scales gleaned for other Kitaev candidates~\cite{Glamazda,Nasu2}.}

{A similar integrated intensity, $I_{\rm low}$, may be obtained on integration over a window $0<\omega<\omega_{\rm min}$ which leads to a scaling of $f(1-f)$~\cite{Nasu2} at low temperatures and is sensitive to the low-energy fermion density of states. In Fig.~\ref{fig3}(c) we plot $I_{\rm low}$ in the low-frequency interval of 10-45 cm$^{-1}$ as obtained from the raw data in a cross-polarised set up and it reveals a general decrease with decreasing temperature. While this is very encouragingly in qualitative agreement with the proposed scaling for the pure Kitaev model~\cite{Nasu2}, our present experimental resolution does not allow us for a quantitative comparison mainly due to the low photon count in the cross-polarised set up used for this experiment. The quantitative analysis of $I_{\rm low}$ is further complicated by the finite low frequency quasi-elastic signal which owes its origin to the dilute disorder present in Cu$_2$IrO$_3$ in addition to small non-Kitaev interactions. Indeed, recent numerical calculations~\cite{PhysRevLett.104.237203,PhysRevB.84.115146,sanyal2020emergent,biswas2020geometry} show that non-magnetic dilution of the honeycomb lattice for the pure Kitaev model can produce a large number of low energy fermionic modes without destroying the essence of the Kitaev QSL. The interaction and the fate of these modes that are clearly relevant for Cu$_2$IrO$_3$ and their contribution to the low energy density of states presently masks the pristine behaviour of the low energy Majorana fermions. However, the lack of magnetic order and the anomalous spin-phonon coupling (see below) along with the mid energy fermionic magnetic continuum clearly shows that the pure Kitaev model and Majorana fermions are right starting point to understand the interplay of fractionalisation and disorder at lower energies along with possible non-Kitaev interactions.}

\begin{figure}[h!]
\centering
\includegraphics[width=8.6cm]{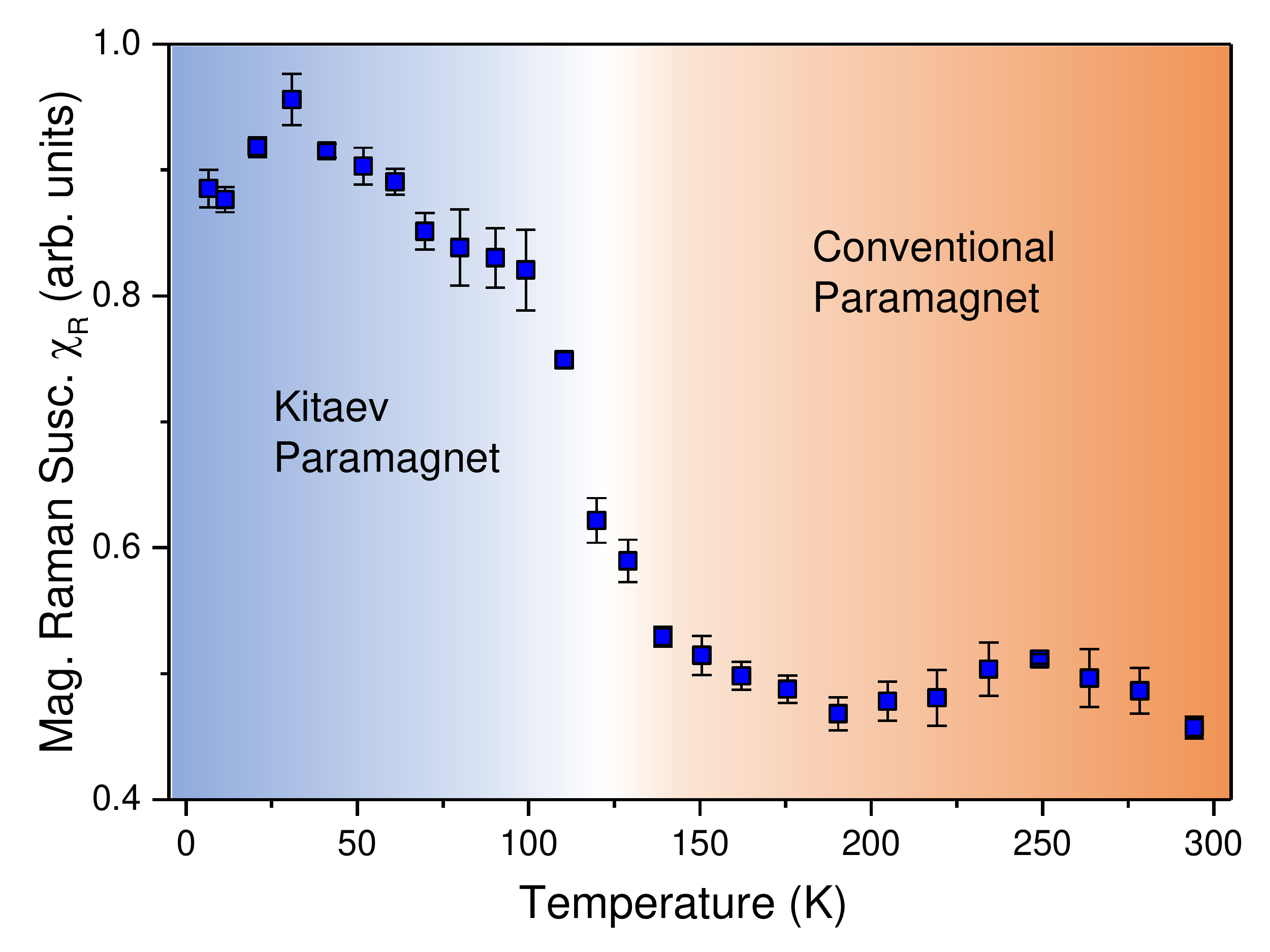}
\caption{\small Magnetic Raman susceptibility as a function of temperature deduced from the Kramers-Kronig relation. The shaded regions mark the boundary between the conventional and Kitaev paramagnetic states.}
\label{fig4}
\end{figure}

{To further extract the Majorana essence from the magnetic background, dynamic spin susceptibility $\chi_R$ is measured. The magnetic Raman susceptibility $\chi_R$ is extracted by integrating Raman conductivity $\frac{\chi"(\omega)}{\omega}$ in the frequency range 70 to 600 cm$^{-1}$ using Kramers-Kronig relation, $\chi_R = \lim_{\omega \to 0} \chi(k=0,\omega) \equiv \frac{2}{\pi} \int\frac{\chi"(\omega)}{\omega} d\omega$. The dynamical Raman tensor susceptibility $\chi"(\omega)$ is proportional to Raman intensity as $I(\omega)=2\pi\int<R(t)R(0)>e^{i\omega t} dt \propto [1+n(\omega)]\chi"(\omega$), where R(t) is the Raman operator. Fig.~\ref{fig4} displays the temperature dependence of $\chi_R$ which shows that it remains almost constant down to 120 K, below which it increases rapidly with decreasing temperature. In the Kitaev QSL state, the Raman operator couples to the dispersing Majorana fermions and extensively projects to the two-Majorana fermion density of states (DOS)~\cite{Perreault}. Thus, an enhancement of $\chi_R$ below 120 K corresponds to significant enhancement of Majorana DOS in the system driving the system from a simple paramagnet to a Kitaev paramagnet, also clearly reflected in temperature dependence of \textit{I$_{mid}$}.}

{Both \textit{I$_{mid}$} in Fig.~\ref{fig3} and $\chi_R$ in Fig.~\ref{fig4} show a subtle decrease below $\sim$25~K. At a first glance, one may correlate this with the partial spin-freezing reported for Cu$_2$IrO$_3$ below $\sim10$~K in recent $\mu$SR and NQR studies~\cite{Kenney,Takahashi}. However, this may not be the case as our samples do not show evidence of spin-freezing down to $2$~K in ac $\chi$ (Appendix~\ref{A_A}) as well as down to $260$~mK in $\mu$SR measurements~\cite{Choi}. It is tempting to associate the decrease in \textit{I$_{mid}$} below 25 K ($\sim0.09J_K$) to the calculated \textit{I$_{mid}$} by quantum Monte Carlo calculations (peaking at $\sim0.07J_K$)~\cite{Nasu2}.}

\section{Majorana-phonon coupling}

{In absence of any thermal phase transition to a magnetic ordered state, the anomalous renormalisation of the phonon frequency and increment in the linewidth at low temperatures suggest that new decay channels are opening up for the phonons to interact and possibly decay into. Given the current understanding of the phenomenology of Cu$_2$IrO$_3$~\cite{Choi,Kenney,Takahashi} and the encouraging match of the the energy-scale $T_h$, it is natural to seek an explanation of the above experiments in terms of the excitations of the Kitaev QSL, {\it i.e.} the Majorana fermions and $Z_2$ fluxes that results from the spin-(optical) phonon coupling. Already, the existing calculations~\cite{Nasu,Nasu2} correctly accounts for the broad magnetic background in Fig.~\ref{fig1} to this end.}

{We now show that the spin-phonon coupling leads to the possibility of a Yukawa-like coupling between a Majorana bilinear and the phonon somewhat akin to the electron-phonon coupling in superconductivity. This coupling, in turn, accounts for the experimental findings and hence provides a very interesting understanding of the experimental data in terms of the Majorana-phonon coupling. Below, we outline our calculations capturing the essence of the above physics. The Kitaev spin model is given by~\cite{Kitaev}}

\begin{align}
    H_{\rm Kitaev}=\sum_{i,\alpha} J_{i,\alpha} S^\alpha_iS_{i+\hat{\alpha}}^\alpha
    \label{eq_ham}
\end{align}

{where $\alpha$ denotes $x,y$ or $z$ type of bonds and $\hat{\alpha}$ denotes the three nearest neighbour vectors of honeycomb lattice. The exchange couplings are functions of the ionic positions as they come from the overlap of the electronic wave-functions. Thus, in presence of Lattice vibrations, we have~\cite{Bhattacharjee}}

\begin{align}
    J_{i,\alpha}=J_K+\frac{\partial J_{i,\alpha}}{\partial R^a_{i,\alpha}}\delta_{i,\alpha}^a+\frac{1}{2}\frac{\partial^2 J_{i,\alpha}}{\partial R_{i,\alpha}^a\partial R_{i,\alpha}^b} \delta^a_{i,\alpha} \delta_{i,\alpha}^b
    \label{eq_spexp}
\end{align}

{where the expansion is done about the equilibrium ionic positions of the crystal, $\bar{R}_{i,\alpha}^a=r^a_i-r^a_{i+\hat{\alpha}}$ with  $\delta_{i,\alpha}^a=R^a_{i,\alpha}-\bar{R}_{i,\alpha}^a$ ($a=x,y$) denoting the deformation of the bond and the derivatives are evaluated at the equilibrium position $\bar{R}_{ij}$.}

{This leads to the spin-phonon Hamiltonian that dictates the coupled dynamics of the optical phonons and the spins}

\begin{align}
    H=H_{\rm spin}+H_{\rm spin-phonon}+H_{\rm phonon}
    \label{eq_hsp_hph_hspin}
\end{align}

{\noindent where $H_{\rm spin}$ is the bare spin Kitaev Hamiltonian of Eq.~\ref{eq_ham} with $J_{i,\alpha}\rightarrow J_K$, $H_{\rm phonon}$ is the bare Harmonic phonon Hamiltonian and}

\begin{align}
    H_{\rm spin-phonon}=H_1+H_2
    \label{eq_hsp}
\end{align}

{\noindent represents the spin-phonon coupling. The two terms denote the first and second order contributions of Eq.~\ref{eq_spexp} and are given by}

\begin{align}
    H_1=\sum_{i,\alpha}\frac{\partial J_{i,\alpha}}{\partial R^a_{i,\alpha}}\delta_{i,\alpha}^a~S^\alpha_iS_{i+\hat{\alpha}}^\alpha 
    \label{eq_ham1}
\end{align}

{\noindent and}

\begin{align}
    H_2=\frac{1}{2}\sum_{i,\alpha} \frac{\partial^2 J_{i,\alpha}}{\partial R_{i,\alpha}^a\partial R_{i,\alpha}^b}\delta^a_{i,\alpha}\delta_{i,\alpha}^b ~S^\alpha_iS_{i+\hat{\alpha}}^\alpha 
    \label{eq_ham2}
\end{align}

{\noindent respectively. Expressing the phonons in terms of their normal modes and neglecting the various form factors which we expect to be unimportant for the generic temperature dependence that we are focussing on, we now obtain the renormalisation of the phonon frequency and linewidth by calculating the self-energy correction to the phonon propagators due to the above spin-phonon interactions within a single mode approximation for the phonons.}

\begin{figure}
\centering
\subcaptionbox{}
       { \begin{tikzpicture}
        \begin{feynman}
        \vertex (a);
        \vertex [right=of a] (b);
        \vertex [above right=of b] (f1);
        \vertex [below right=of b] (f2);
    
        \diagram* {
        (a) -- [boson, edge label'=\({\bf k}~\Omega\)] (b) -- [fermion, edge label'=\({\bf k-k'}~\Omega-\omega\)] (f1),
        (b) -- [fermion, edge label'=\({\bf k'}~\omega\)] (f2),
        };
        \end{feynman}
        \end{tikzpicture}}
    \subcaptionbox{}
        {\begin{tikzpicture}
        \begin{feynman}
        \vertex (a) ;
        \vertex [right=of a] (b);
        \vertex [above right=of b] (f1);
        \vertex [below right=of b] (f2);
    
        \diagram* {
        (a) -- [boson, edge label'=\({\bf k}~\Omega\)] (b) -- [fermion, edge label=\({\bf k'}~\omega\)] (f1),
        (b) -- [anti fermion, edge label=\({\bf k'-k}~\Omega-\omega\)] (f2),
        };
        \end{feynman}
        \end{tikzpicture}}
\subcaptionbox{}
        {\begin{tikzpicture}
        \begin{feynman}
        \vertex (a) ;
        \vertex [below right=of a] (b);
        \vertex [below left=of b] (c);
        \vertex [above right=of b] (f1);
        \vertex [below right=of b] (f2);
    
        \diagram* {
        (a) -- [boson, edge label'=\({\bf k}\)] (b) -- [fermion, edge label'=\({\bf k+k'-k''}\)] (f1),
        (c) -- [boson, edge label=\({\bf k'}\)] (b)--[fermion, edge label'=\({\bf k''}\)] (f2),
        };
        \end{feynman}
        \end{tikzpicture}}
    \subcaptionbox{}
        {\begin{tikzpicture}
        \begin{feynman}
        \vertex (a) ;
        \vertex [below right=of a] (b);
        \vertex [below left=of b] (c);
        \vertex [above right=of b] (f1);
        \vertex [below right=of b] (f2);
    
        \diagram* {
        (a) -- [boson, edge label'=\({\bf k}\)] (b) -- [fermion, edge label=\({\bf k''}\)] (f1),
        (c) -- [boson, edge label=\({\bf k'}\)] (b)--[anti fermion, edge label=\({\bf k''-k-k'}\)] (f2),
        };
        \end{feynman}
        \end{tikzpicture}}
    \caption{\small ({\bf a}), ({\bf b}) and ({\bf c}), ({\bf d}) denote the interaction between phonon and matter fermion~\cite{Knolle2} coming from the first and second order contributions to the spin-phonon coupling, respectively.}
    \label{fig:spinphonon}
\end{figure}
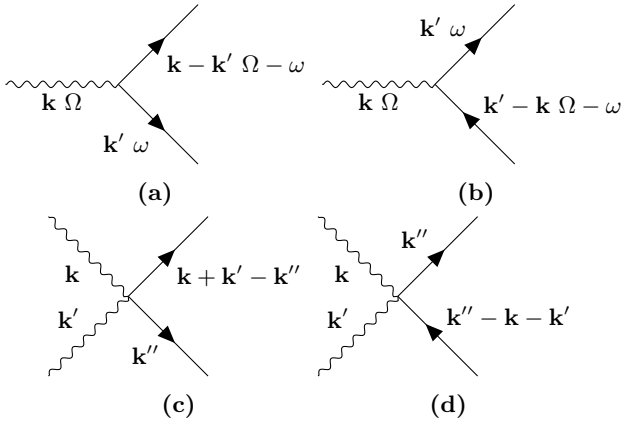

{Within the Kitaev QSL phenomenology, we perform the standard Majorana decoupling of the spins to obtain (in the zero-flux sector) the scattering vertices between the matter fermions and the phonons as shown in Fig.~\ref{fig:spinphonon}. Here we have performed the well-known~\cite{Knolle2} transformation of the Majoranas to the bond matter fermions for the Kitaev QSL. Also, the upper and lower panels denote the interaction vertices arising due to $H_1$ (Eq.~\ref{eq_ham1}) and $H_2$ (Eq.~\ref{eq_ham2}), respectively.}

{These interactions clearly show that the phonon can decay into the fractionalised excitations of the QSL and this would renormalise both the frequency and the linewidth of the phonon peak. In regard to the linewidth, we expect that an anomalous broadening as the temperature is decreased since on lowering the temperature the fermions become more coherent and hence the phonon can more efficiently decay into them while obeying all the conservation laws.}

\subsection{Frequency renormalisation}

{The leading order contribution to the renormalisation of the frequency comes from the lower panel of Fig.~\ref{fig:spinphonon} when we integrate over the fermions. From Eq.~\ref{eq_ham2}, the resultant frequency renormalisation is given by}

\begin{align}
    \delta\omega\propto\frac{1}{N_b}\sum_{i,\alpha}\langle \bar J_K~S^\alpha_iS^\alpha_{i+\hat{\alpha}}\rangle_S 
    \label{eq_shift}
\end{align}

{\noindent where $\langle\cdots\rangle_S$ denotes averaging of the equal time spin correlators over the thermodynamic ensemble and the proportionality constant is given in terms of the first order spin-phonon coupling and the transformation to the phonon soft modes. For the present discussion we neglect their detail structure and assume it to be a constant, $\lambda$.}

{Within a free Majorana phenomenology, the spin-energy can be calculated in the zero-flux sector to obtain an estimate of $\delta\omega$. This calculation is detailed in Appendix~\ref{A_I} and it readily matches the expectation that it goes to zero at $T\rightarrow\infty$ and gradually turns non-zero around $T \sim J_K$ ultimately saturating to a negative constant number at zero temperature corresponding to the ground state energy-density of the spins. Further, numerical calculations exist for finite temperatures including all the flux sectors for the pure Kitaev model~\cite{Nasu} which shows that a rather sharp crossover from zero to non-zero values. With the energy being generally negative this nominally suggests softening of the phonon frequency. We however note that the mode dependence of the above contribution is entirely due to the matrix elements which we have neglected in this calculation. Further temperature dependence can come from the real part of the self-energy of the bubble (see Appendix~\ref{A_I} for details).}

\subsection{Phonon Linewidth}

{The leading contribution to the linewidth, however comes from the bubble contributions arising due to the two vertices in the lower panel of Fig.~\ref{fig:spinphonon}. At finite temperature and in presence of spin-interactions beyond the pure Kitaev model, clearly the fermion lines would be further renormalised by its scattering with the $Z_2$ fluxes, which, in turn, provide finite lifetime to the fermions as well as renormalise their bandwidth~\cite{Kitaev}. For very low temperature and within the exactly solvable model, we neglect the scattering with the gapped $Z_2$ fluxes and then we have free Majorana fermions which seems to be justified on the basis of the numerical calculations~\cite{Nasu,Nasu2} which shows that the qualitative features of the matter fermion density of states remain intact at finite temperatures almost all the way up to $T_h$. Within this free Majorana phenomenology, we now calculate the leading order contribution to the Raman linewidth computing the self-energy bubble diagram for a particular normal mode. The  imaginary part of the phonon self-energy correction at the leading order is then given by}

\begin{align}
    Im[\Sigma({\bf q},&\omega+i0^+)]\propto\frac{J_K^2}{N_b}\sum_{\bf k}(1-n_F(\epsilon_{\bf k})-n_F(\epsilon_{\bf k+q}))\nonumber\\
    &\times\Big[\delta(\omega+\epsilon_{\bf k}+\epsilon_{\bf k+q})\Big.\Big.-\delta(\omega-\epsilon_{\bf k}-\epsilon_{\bf k+q})\Big]
    \label{eq_imself}
\end{align}

{\noindent where, again, the proportionality constant depends on the magneto-elastic coupling and the normal-mode matrix elements which have been assumed to be a constant($\chi$) for this calculation. Here $n_F(\epsilon_{\bf k})$ denotes the fermion occupancy of the complex fermionic modes with dispersion $\epsilon_{\bf k}$ in the zero-flux sector. This contribution, as the delta function indicates, arises due to the decay of the phonon into two fermions. As $T\rightarrow\infty$, the Majorana fermions become incoherent and hence the above contribution to linewidth goes to zero, while at low temperatures it reaches a finite value for the completely coherent Majorana fermions.} 

{This effect is completely opposite of the usual temperature related broadening due to anharmonic terms and arises due to the development of a coherent scattering channel for the phonons. Clearly, in absence of any magnetic phase transition, such coherent particles \-- in case of a Kitaev QSL Majorana fermions \-- indicate novel low temperature physics in the spin sector. This is in direct conformity with the experimental observation. Once the flux excitation is taken into account, it only renormalises the free Majorana contribution without changing its qualitative features. We note that the real part of the self-energy coming from the bubble further renormalises the phonon frequency and hence contributes to $\delta\omega$ in Eq. \ref{eq_shift}. Here we neglect such higher order contributions.}

{The phonon intensity obtained from the above calculation is given by the Lorentzian form}

\begin{align}
    %Im[D({\bf q},\omega+i0^+)]=
    \frac{4\omega_0^2 Im[\Sigma]}{(\omega^2-\omega_0^2-2\omega_0\delta\omega)^2+4\omega_0^2Im[\Sigma]^2}
\end{align}

{\noindent where $\omega_0$ is the bare phonon frequency of a particular normal mode. We evaluate the above expression considering $q\rightarrow0$ limit which is relevant to the experiment (see Eq.~\ref{eq_imself} of Appendix~\ref{A_I}).} 

{We plot the Stokes line in Fig.~\ref{fig:lorentzian}. This is in qualitative agreement with the experimental data. A further comparison with the experimental data is obtained by fitting the results to the experimental data as shown in Appendix~\ref{A_I}.} 

\begin{figure}
    \centering
    \includegraphics[width=8.6cm]{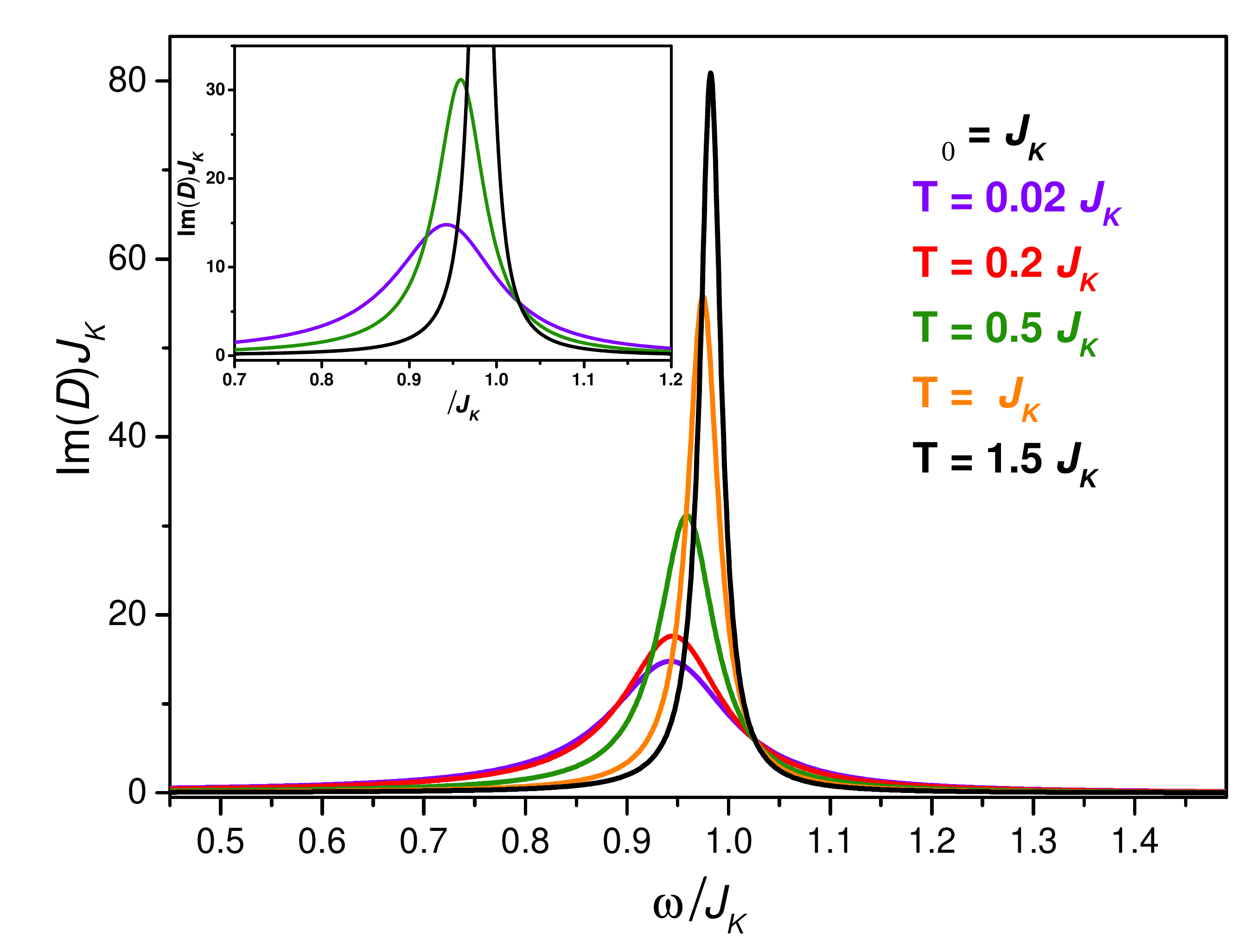}
    \caption{\small Imaginary part of the phonon Green's function ($D$) with the frequency scaled w.r.t. the Kitaev coupling $J_K$. Different curves represent different temperatures. The spin-phonon coupling constants are taken as $\lambda=0.13$ and $\chi=0.2$. }
    \label{fig:lorentzian}
\end{figure}

{To account for the temperature dependence of intensity for the $M3$ mode, we note that the intensity is generically of the form~\cite{suzuki1973theory} $|A+B_\alpha\langle S^\alpha_i S^\alpha_{i+\hat{\alpha}}\rangle|^2$, {\it i.e.} proportional to the nearest neighbour spin correlations. This, within free Majorana fermions, can be calculated to yield a dependence proportional to $|\delta\omega|$ as is clear from Eq.~\ref{eq_ham}. While such effects should appear for all the four modes, the particular sensitivity of $M3$ appears to us as a matrix-element effect that requires more detailed calculations.}

\section{Summary and Outlook}

{To summarise, we have investigated the Raman response of the ``second-generation" Kitaev QSL candidate Cu$_2$IrO$_3$. In addition to the the magnetic continuum (consistent with the Kitaev coupling, $J_K\approx 24$ meV) observed in the ``first-generation" Kitaev materials, we observe clear anomalous renormalisation of the Raman-active phonons below $\sim$120 K. Encouraged by the conformity of the energy-scales of the magnetic continuum and the phonon anomaly within a Kitaev phenomenology, we investigate the qualitative features of the Majorana-(optical) phonon coupling to make an estimate for the phonon anomaly which accounts for the experimental observations.} 

{Our results thus provide strong indication for the relevance of Kitaev QSL physics and the immensely exciting possibility of positive identification of fractionalisation in the nearly perfect honeycomb iridate Cu$_2$IrO$_3$. The phonon anomalies below a characteristic temperature provide yet another Raman signature of fractionalized Majorana fermions in addition to the magnetic continuum in prospective candidates of QSL. Although our samples have a smaller amount of Cu$^{2+}$ impurities, as evidenced by the absence of a spin-glass transition or phase-separation into QSL and magnetically frozen volume fractions~\cite{Choi}, we do find that the residual $\sim 5\%$ Cu$^{2+}$ impurities lead to a random-singlet phase below $T \sim$10-20~K. This small amount of disorder most likely also leads to the quasi-elastic Raman signal at low frequencies. This opens several interesting questions from both experimental and theoretical sides. On the experimental front, several future directions of study can be envisaged. Single crystals of Cu$_2$IrO$_3$ are not currently available. With high quality single crystals, the intrinsic low frequency Raman signal can be revealed and compared with expectations for the pure Kitaev model. Additionally, with single crystals, polarisation dependent Raman studies will become possible which will allow a further quantative comparison between theoretical calculations and experiments to further substantiate the physics of Majorana-phonon coupling. Further, inelastic neutron scattering to measure the energy and momentum dependent excitation spectrum are desirable to be able to make quantitative comparisons with specific Hamiltonians including the Kitaev model and its extensions. Finally, crystals will allow looking for quantisation in thermal Hall measurements, similar to what has been reported for $\alpha$-RuCl$_3$~\cite{Matsuda}, although recent experimental works have shown that the field induced paramagnetic state in RuCl$_3$ may not be a QSL afterall~\cite{Ponomaryov,Bachus}. On the theoretical side, the present calculation only takes care of the free Majorana fermions while neglects the fluxes as well as other non-Kitaev spin interactions. Their roles in the present calculations needs to be quantitatively settled both for Cu$_2$IrO$_3$ and other QSLs in general to investigate the physics of fractionalisation through phonons. Since the appearance of our work (our arXiv reference~\cite{Pal}), other theoretical studies on spin-phonon coupling in a Kitaev QSL has recently been developed in Ref.~[\onlinecite{Metavitsiadis2}-\onlinecite{Yang}] which can further account for quantitative features of the vibrational Raman spectra in a Kitaev QSL beyond universal temperature dependence as attempted here. Such a quantitative treatment along would require a more detailed knowledge of Hamiltonian and phonon parameters for Cu$_2$IrO$_3$ that is presently missing and forms an important future direction.}

\begin{acknowledgments}

{AKS thanks Nanomission Council and the Year of Science professorship of DST for financial support. PS, AA and YS thank the X-ray, liquid Helium plant and the SQUID magnetometer facilities at IISER Mohali. SB acknowledges J. Knolle and R. Moessner for previous collaborations and A. Nanda and K. Damle for discussions. SB acknowledge Max Planck Partner Grant at ICTS and SERB-DST (India) project grant No. ECR/2017/000504 for funding. SB and AS acknowledge support of the Department of Atomic Energy, Government of India, under project no.12-R$\&$D-TFR-5.10-1100.}

\end{acknowledgments}

\appendix

\section{Materials and Magnetic characterisation}
\label{A_A}

{High quality polycrystalline samples of Cu$_2$IrO$_3$ were synthesised by an ion exchange reaction by mixing Na$_2$IrO$_3$ and CuCl in the mole ratio $1 : 2.05$~\cite{Abramchuk}.  The mixture with total mass $350$~mg was pelletised, placed in an alumina crucible and sealed under vacuum in a quartz tube. The tube was heated at $1^o$C/min to $350^o$C, kept at that temperature for $16$~h and then cooled to room temperature at the same rate. Then the product was ground into a fine powder and washed five times with ammonium hydroxide (NH$_4$OH) and twice with distilled water. After being washed, the resulting material was dried at room temperature under vacuum for $2$~h.}

\begin{figure}[h!]
\centering
\includegraphics[width=8.6cm]{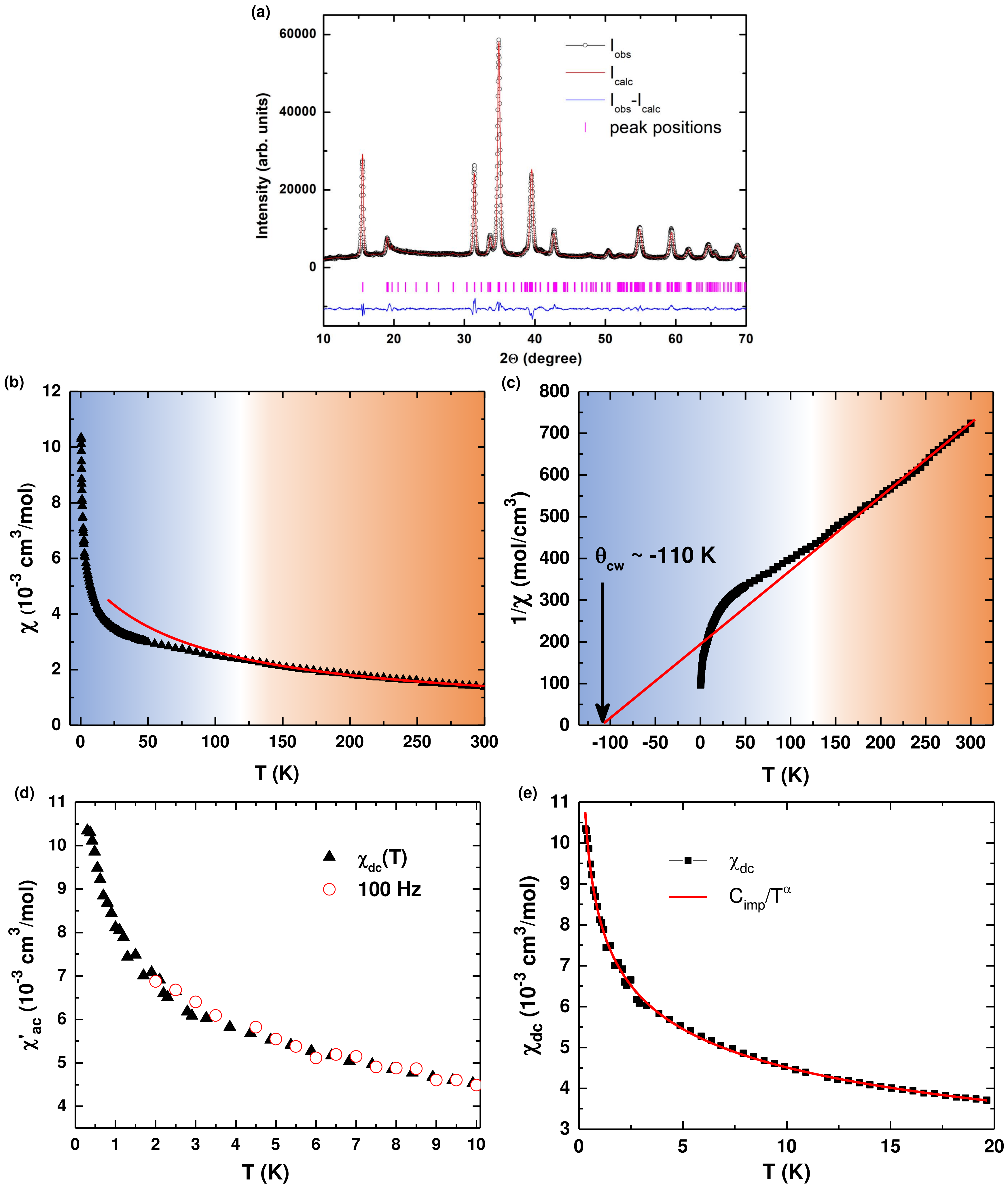}
\caption{\small ({\bf a}) Powder x-ray diffraction pattern for Cu$_2$IrO$_3$ and its Rietveld refinement. ({\bf b}) Temperature dependence of the dc magnetic susceptibility between $0.3$ and $300$~K\@. The red solid line is the Curie-Weiss (CW) fit over the range 120-300 K. ({\bf c}) Inverse susceptibility fitted to CW form (solid red line) over the range 120-300 K. The shaded regions demarcate the boundary between conventional and Kitaev paramagnet where $\chi_{DC}$ starts deviating from the CW behaviour.({\bf d}) Temperature variation of ac susceptibility down to 2 K in addition to the dc susceptibility down to 0.3 K. ({\bf e}) Low temperature $\chi_{dc}$ data fit to a sub-Curie law.}
\label{fig_a0}
\end{figure}

{The powder x-ray diffraction and a Rietveld refinement of the same is shown in Fig.~\ref{fig_a0}(a) confirming that a single-phase product with the correct crystallographic structure (monoclinic \textit{C2/c}) is obtained. The dc magnetic susceptibility between $0.3$ and $300$~K is shown in Fig.~\ref{fig_a0}(b). The high temperature $\chi$ in the region $120 < T < 300~K$ is fit well with the Curie-Weiss (CW) form $\chi = \chi_0 + C/(T-\Theta_{CW})$ giving an effective moment close to $S=1/2$ and $\Theta_{CW} \approx$ -100~\textit{K}. Fig.~\ref{fig_a0}(c) shows the CW behaviour of the inverse DC susceptibility fitted in the range of 120 to 300 K. The fit is extrapolated to extract the Curie-Weiss temperature $\Theta_{CW} \approx$ - 110~\textit{K}. The deviation of susceptibility from the CW law below the Majorana crossover temperature \textit{T$_h$} is similar to that observed in $\alpha$-RuCl$_3$~\cite{Do} and in the Quantum Monte Carlo calculations by Nasu et al.~\cite{Nasu}. Below $10$~K, the $1/\chi$ data shows a sharp downturn, which is the contribution from disorder, the magnitude of which we estimate below. The $\chi_{\rm dc}$ data below 10 K is shown in Fig.~\ref{fig_a0}(d). We did not observe any signatures of spin-glass freezing reported previously~\cite{Abramchuk}. Absence of freezing is confirmed by an ac susceptibility measurement down to 2 K which is also shown in Fig.~\ref{fig_a0}(d). Fig.~\ref{fig_a0}(e) shows that the low temperature $\chi$ data follows a sub-Curie law behaviour $\chi = C_{imp}/T^\alpha$ with $C_{imp} \approx 0.019(3)$~cm$^3$~K/mol and $\alpha = 0.23(2)$. This $T$ dependence is consistent with a Random-singlet state. However, the magnitude of $C_{imp}$ gives an $\approx 5\%$ estimate for the fraction of impurity spins which participate in this low temperature random-singlet state.  This impurity concentration is roughly half of the previously reported samples for which a spin-glass state was observed in magnetic measurement~\cite{Kenney}.}

\section{Choice of $\omega_{min}$ and $\omega_{max}$ in mid-frequency background}
\label{A_B}

\begin{figure}[h!]
\centering
\includegraphics[width=8.6cm]{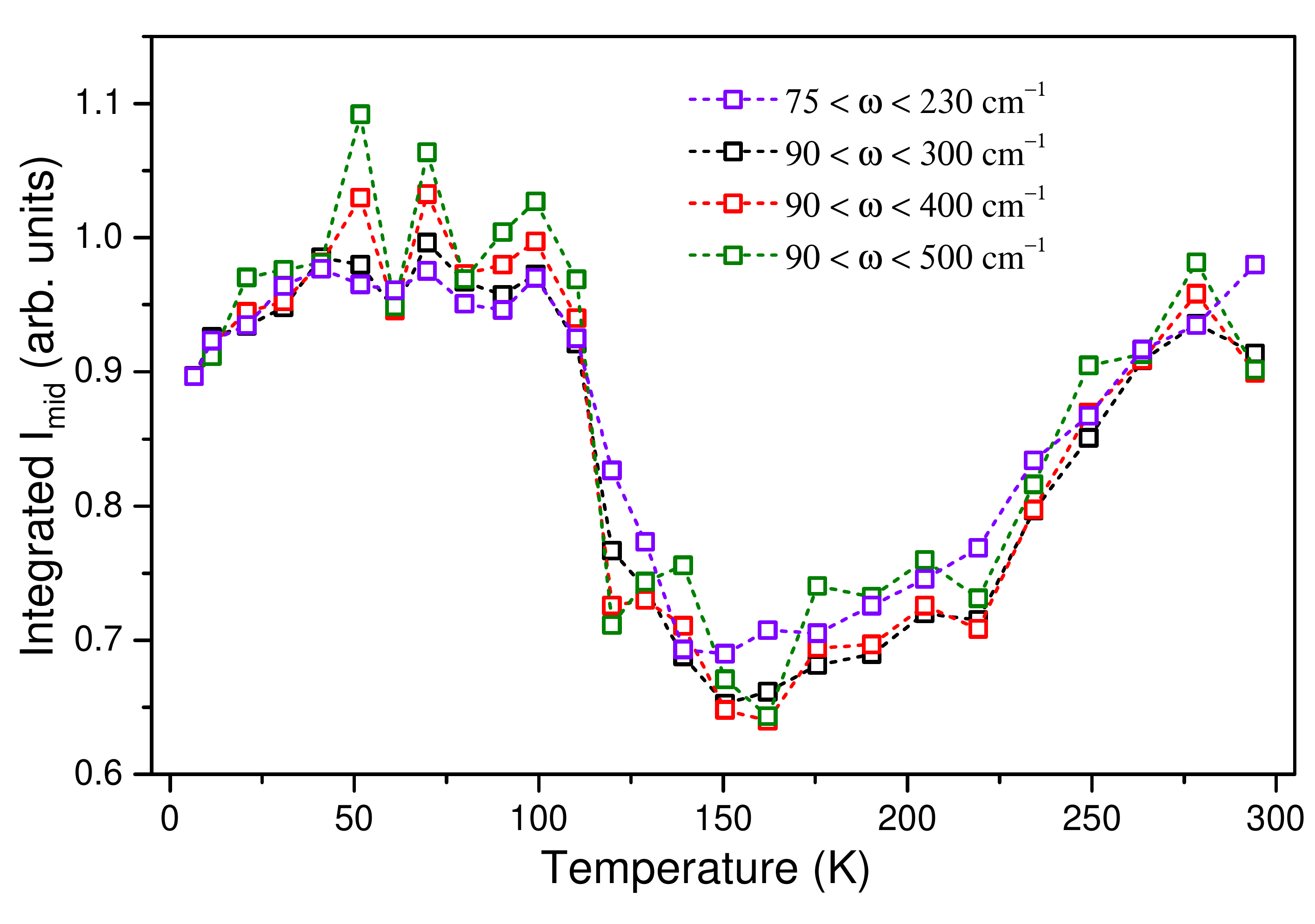}
\caption{\small Temperature dependence of integrated background intensity (normalised) for different energy ranges.}
\label{fig_a1}
\end{figure}

{Theoretical predictions on choice of $\omega_{min}$ and $\omega_{max}$ suggest the energy range of $0.5 < \omega/J < 1.25$~\cite{Nasu2} which gives a frequency window of $97 < \omega < 242$ cm$^{-1}$ for Cu$_2$IrO$_3$ taking \textit{J$_K$} = 24 meV. But still there is no strong foundation on selection of this intermediate energy range and various scales have been chosen for different Kitaev materials. Such as, for Li$_2$IrO$_3$, $I_{mid}$ was chosen to be $1.5 < \omega/J < 3$~\cite{Glamazda}, whereas for $\alpha$-RuCl$_3$, an energy range of $0.6 < \omega/J < 1.9$ was adopted by the authors~\cite{Glamazda2}. To inspect the robustness for the $I_{mid}$ range selection for Cu$_2$IrO$_3$, we calculated the integrated areas of the background taking different $\omega$ ranges, and the results are plotted in Fig.~\ref{fig_a1}. We find that the scaling behaviour of $I_{mid}$ is the same for these moderate variations of the window size. Range of $120 < \omega < 260$ cm$^{-1}$ is chosen for Cu$_2$IrO$_3$ since it is least affected by the interference of stong phonon intensities.}

\section{Different Bosonic and fermionic fits to integrated \textit{I}$_{mid}$}
\label{A_C}

\begin{figure}[h!]
\centering
\includegraphics[width=8.6cm]{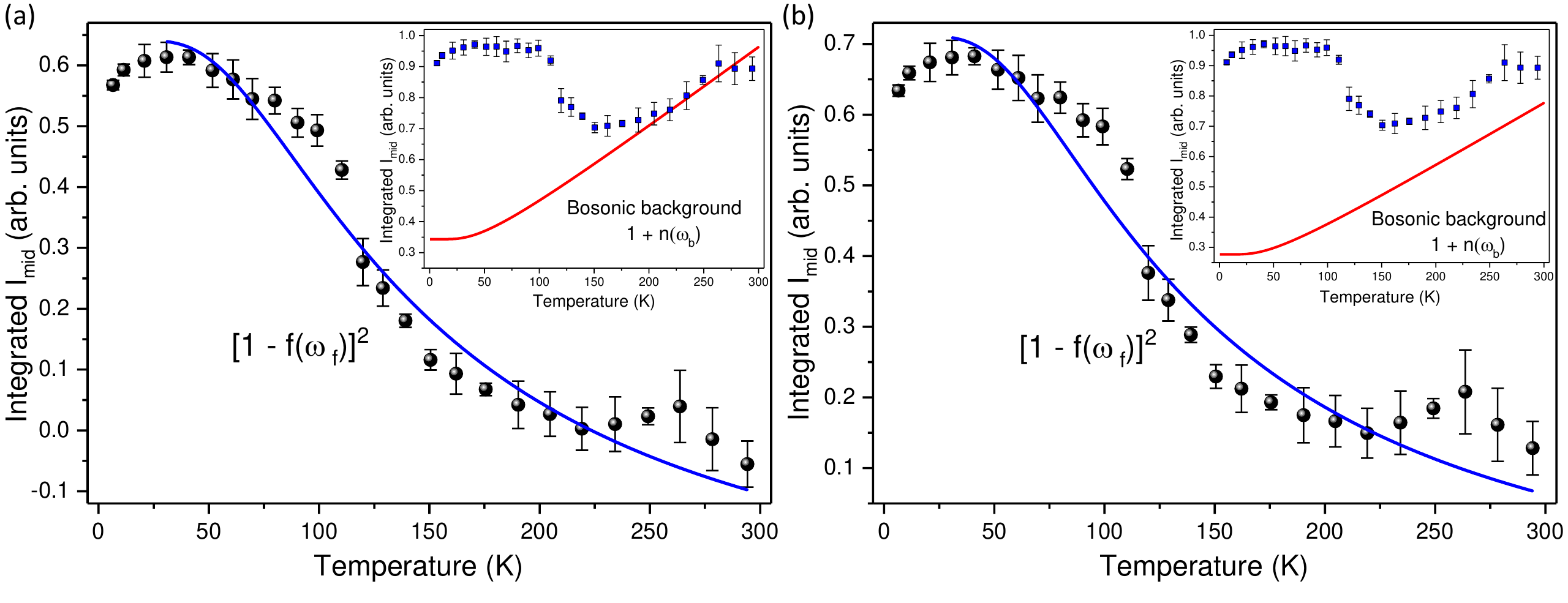}
\caption{\small ({\bf a})-({\bf b}) Fermionic fits to integrated \textit{I$_{mid}$} with two different Bosonic backgrounds (shown in the insets).}
\label{fig_a2}
\end{figure}

Fig.~\ref{fig_a2}(a)-(b) shows variation of the fermionic fit to the integrated intensity in the frequency range 120 $< \omega <$ 260 cm$^{-1}$ taking small variations in the Bosonic fits. The fitting parameter $\omega_f$  deviate in the first decimal place compared to the fit shown in Fig.~\ref{fig3} and hence the Fermionic fit is robust under the modulations done in the Bosonic background. The fitting in Fig.~\ref{fig3} is considered due to lower errors in the fitting parameters.

\section{Phonon fits}
\label{A_D}

\begin{figure}[h!]
\centering
\includegraphics[width=8.6cm]{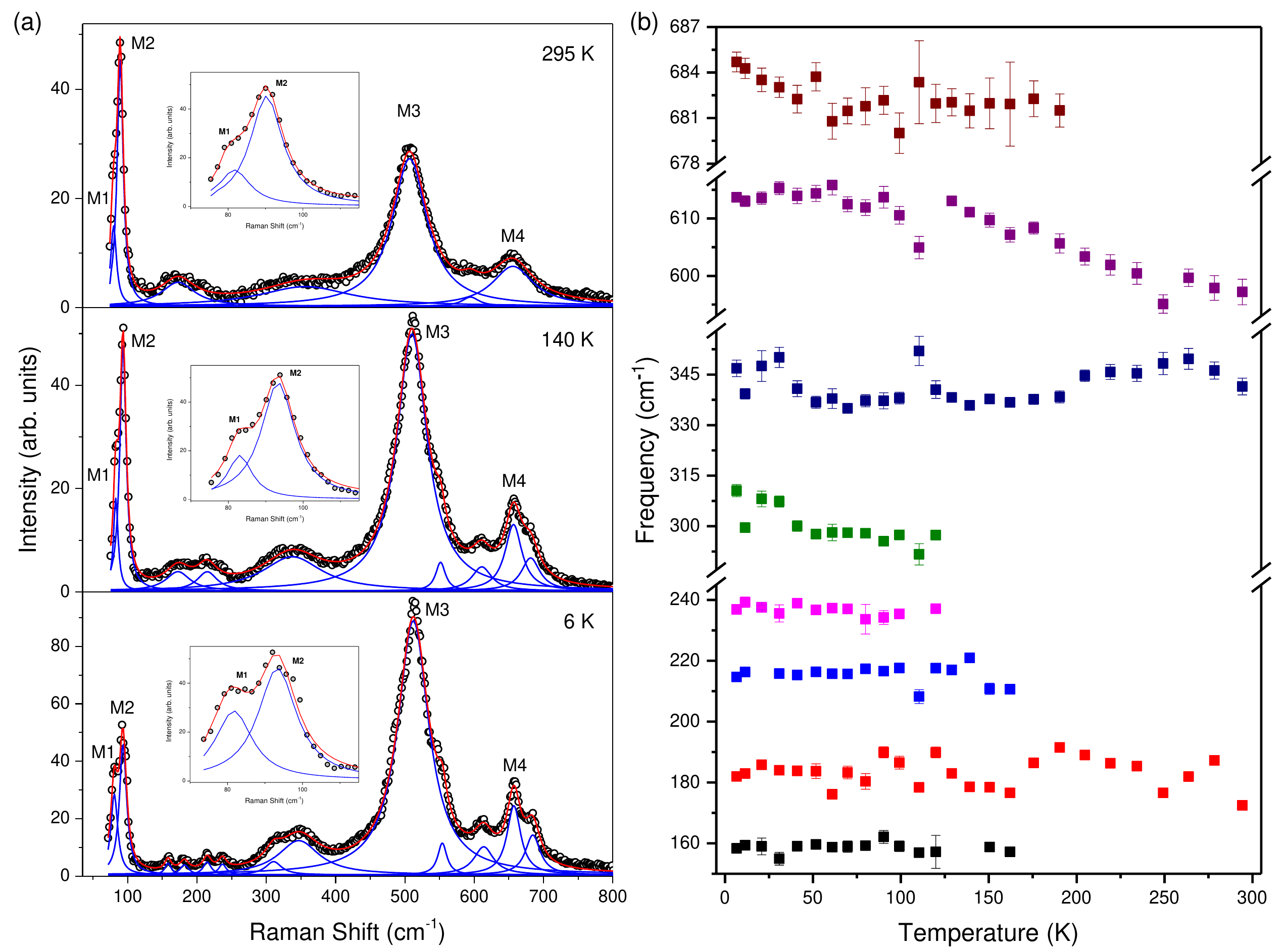}
\caption{\small ({\bf a}) Phonon fits (after subtracting the low-frequency background) to the Raman spectra of Cu$_2$IrO$_3$ at selected temperatures. Experimental data are indicated by black circles. Blue and red curves are individual phonon modes and the cumulative fits, respectively. The magnified M1, M2 doublets are shown in the insets. ({\bf b}) Temperature evolution of weaker phonon modes apart from M1, M2, M3 and M4.}
\label{fig_a3}
\end{figure}

{Fig.~\ref{fig_a3}(a) represents fitted Raman spectra at selected temperatures with the blue curves showing Lorentzian line-shaped phonon modes. The temperature evolution of the weaker and broader phonon frequencies is depicted in Fig.~\ref{fig_a3}(b).}

\section{Lattice anharmonicity}
\label{A_E}

{The impact of changing temperature on phonon population is well described under intrinsic anharmonic effects. Restricting to cubic corrections to phonon self-energy where a phonon decays into a pair of two phonons conserving energy and momenta, the phonon frequency and FWHM (real and imaginary parts of phonon self-energy, respectively) can be given as~\cite{Klemens},

\begin{equation}
\omega(T)= \omega_0 + A[1+2n(\frac{\omega_0}{2})]
\end{equation}

\begin{equation}
\Gamma(T)= \Gamma_0 + B[1+2n(\frac{\omega_0}{2})]
\end{equation}

where, $\omega_0$ and $\Gamma_0$ are frequencies and linewidths at absolute zero, A (negative) and B (positive) are constants, and $n(\frac{\omega_0}{2})$ is the Bose-Einstein thermal factor. In the fits shown in the main text, $\omega_0$ is extracted from the frequency fits in the high-temperature region (120-295 K) and those values of $\omega_0$ are used to fit the FWHMs. The values for the fitting parameters $\omega_0$, $\Gamma_0$, A, and B for different modes are shown in Table~\ref{table1} below.}

\begin{table}[!ht]
\begin{center}
\caption{\small List of fitting parameters for the cubic anharmonic fits to the phonon modes of Cu$_2$IrO$_3$}
    \begin{tabularx}{\columnwidth}{X X X X X}
    \hline \hline \\ [-1.5ex]
    Mode & $\omega_0$ & A & $\Gamma_0$  & B \\ [1ex]
    \hline \\ [-1.5ex]
    M1 & 83.6 & - 0.09 & 5.9 & 0 \\
    M2 & 95.3 & - 0.5 & 9.6 & 0 \\
    M3 & 518.1 & - 6.6 & 41.4 & 13.5 \\
    M4 & 659.9 & - 2.4 & -58.3 & 86.54 \\
    \hline \hline
    \end{tabularx}
    \label{table1}
    \end{center}
\end{table}

\section{Integrated intensity of high-frequency M3 mode}
\label{A_F}

\begin{figure}[h!]
\centering
\includegraphics[width=8.6cm]{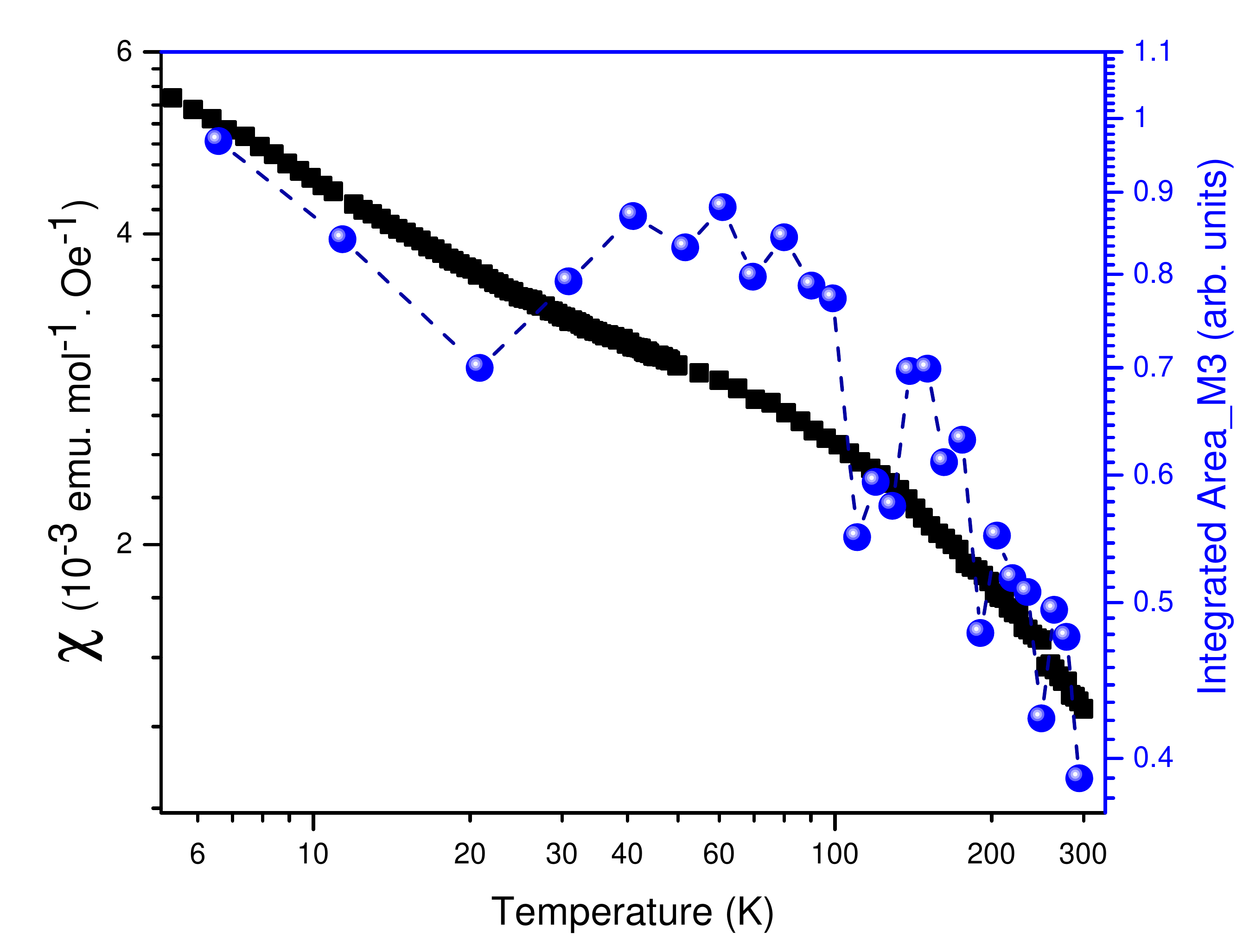}
\caption{\small Temperature variation of integrated intensity of the M3 mode, similar to that of DC magnetic susceptibility, both plotted in log-log scale.}
\label{fig_a4}
\end{figure}

{Fig.~\ref{fig_a4} shows temperature variation of integrated intensity the 510 cm$^{-1}$ Raman mode (M3) of Cu$_2$IrO$_3$ along with DC magnetic susceptibility. Both deviate from their high-temperature behaviour below $\sim$120 K following modulation in the spin correlations in the Kitaev paramagnetic phase.}

\section{Details of the spin-phonon coupling}
\label{A_G}

\subsection{Details of the phonon Hamiltonian and the single mode approximation}
The bare harmonic phonon Hamiltonian is given by,

\begin{align}
    H_{\rm phonon}=\sum_{i,\alpha}\frac{{\bf P}_{i,\alpha}^2}{2m}+\frac{1}{2}\sum_{i,\alpha}\mathcal{D}_{\alpha\beta}^{ab}\delta_{i,\alpha}^a\delta_{i,\beta}^b
    \label{eq_hph}
\end{align}

We can go to the normal mode basis ($\boldsymbol{\Theta}_{i,\bar{p}}$) of phonons by an unitary rotation of $\boldsymbol{\delta}_{i,\alpha}$ which diagonalizes the matrix $\mathcal{D}^{ab}_{\alpha\beta}$.

\begin{align}
    \delta^a_{i,\alpha}=\Gamma^{ab}_{\alpha \bar{p}}\Theta^b_{i,\bar{p}}
\end{align}

In the following sections, we will do a single mode approximation and consider only a particular mode (say $\bar{p}$) to calculate its frequency shift and linewidth broadening. With this assumption, we neglect the possible coupling between the normal modes through the spin-phonon interaction.

\subsection{The spin-(optical) phonon Hamiltonian}

We rewrite the $H_{spin-phonon}$(Eq.~\ref{eq_hsp}) in terms of the normal modes.

\begin{align}
    H_1=\sum_{i,\alpha}\frac{\partial J_{i,\alpha}}{\partial R^a_{i,\alpha}}\Gamma^{ab}_{\alpha\bar{p}}\Theta^b_{i,\bar{p}}~S^\alpha_iS_{i+\hat{\alpha}}^\alpha 
\end{align}
\begin{align}
     H_2=\frac{1}{2}\sum_{i,\alpha} \frac{\partial^2 J_{i,\alpha}}{\partial R_{i,\alpha}^a\partial R_{i,\alpha}^b}\Gamma^{ac}_{\alpha\bar{p}}\Gamma^{bd}_{\alpha\bar{p}}\Theta^c_{i,\bar{p}}\Theta_{i,\bar{p}}^d ~S^\alpha_iS_{i+\hat{\alpha}}^\alpha 
\end{align}

Here, the index $\bar{p}$ is not summed over. Usually, due to overlap of the orbitals, in insulators

\begin{align}
    J_{i,\alpha}= J_K e^{-\eta\delta_{i,\alpha}}
    \label{eq_Jij}
\end{align}

where we have assumed a simplified isotropic form where $\eta$ is the inverse length-scale of decay of overlap. Using the above form, we further define following notations for the compactness of the calculation:

\begin{align}
    \Gamma^{ab}_{\alpha\bar{p}}\frac{\partial}{\partial R^a_{i,\alpha}}e^{-\eta\delta_{i,\alpha}}\equiv\chi^b_{\bar{p},\alpha}\indent\nonumber\\
    \frac{1}{2}\Gamma_{\alpha\bar{p}}^{ac}\Gamma_{\alpha\bar{p}}^{bd}\frac{\partial^2}{\partial R^a_{i,\alpha}\partial R^b_{i,\alpha}}e^{-\eta \delta_{i,\alpha}}\equiv \lambda^{cd}_{\bar{p},\alpha}
\end{align}

\section{Details of the Majorana fermion- (optical) phonon coupling}
\label{A_H}

\noindent In this section, we rewrite the $H_{spin-phonon}$ in terms of fractionalised Majorana degrees of freedom. We follow the standard Kitaev formulation~\cite{Kitaev} to write the spin as,

\begin{align}
    S^\alpha_i=\frac{i}{2}b^\alpha_i c_i
\end{align}

where $b^\alpha_i$ and $c_i$ are the four Majorana fermions. Therefore, we have

\begin{align}
    S^\alpha_iS^\alpha_{i+\hat\alpha}=-\frac{1}{4}u^\alpha_{i,i+\hat\alpha}(ic_ic_{i+\hat\alpha})
\end{align}

Here, $u^\alpha_{i,i+\hat{\alpha}}=ib^\alpha_ib^\alpha_{i+\hat{\alpha}}$. The following calculations will be restricted to the zero-flux sector of the $Z_2$ connection, where we have $u^\alpha_{i,i+\hat{\alpha}}=+1$.

\subsection{The linear term \texorpdfstring{$H_1$}{}}

Using the above transformation, we write down the Eq.~\ref{eq_ham1} in terms of Majorana fermions.

\begin{align}
    H_1=-\frac{ J_K}{4}\sum_{i,\alpha} \chi^a_{\bar{p},\alpha}\Theta_{i,{\bar{p}}}^a(ic_ic_{i+\hat\alpha})
\end{align}

As mentioned in the main text, we now ignore all the matrix element effects and assume, $\chi_{\bar{p},\alpha}^a\equiv \chi^a$. This assumption only changes the vertex functions of the Feynman diagram. It doesn't really change the nature of the virtual processes involved which contributes to self energy of the phonon. For convenience of calculations, we convert the Majorana operators to complex fermion operators (bond matter fermion) within each unit cell (we take two sites joined by a $z$ bond as the unit cell) as~\cite{Knolle2}

\begin{figure}[!h]
    \centering
    \includegraphics[width=5.6cm]{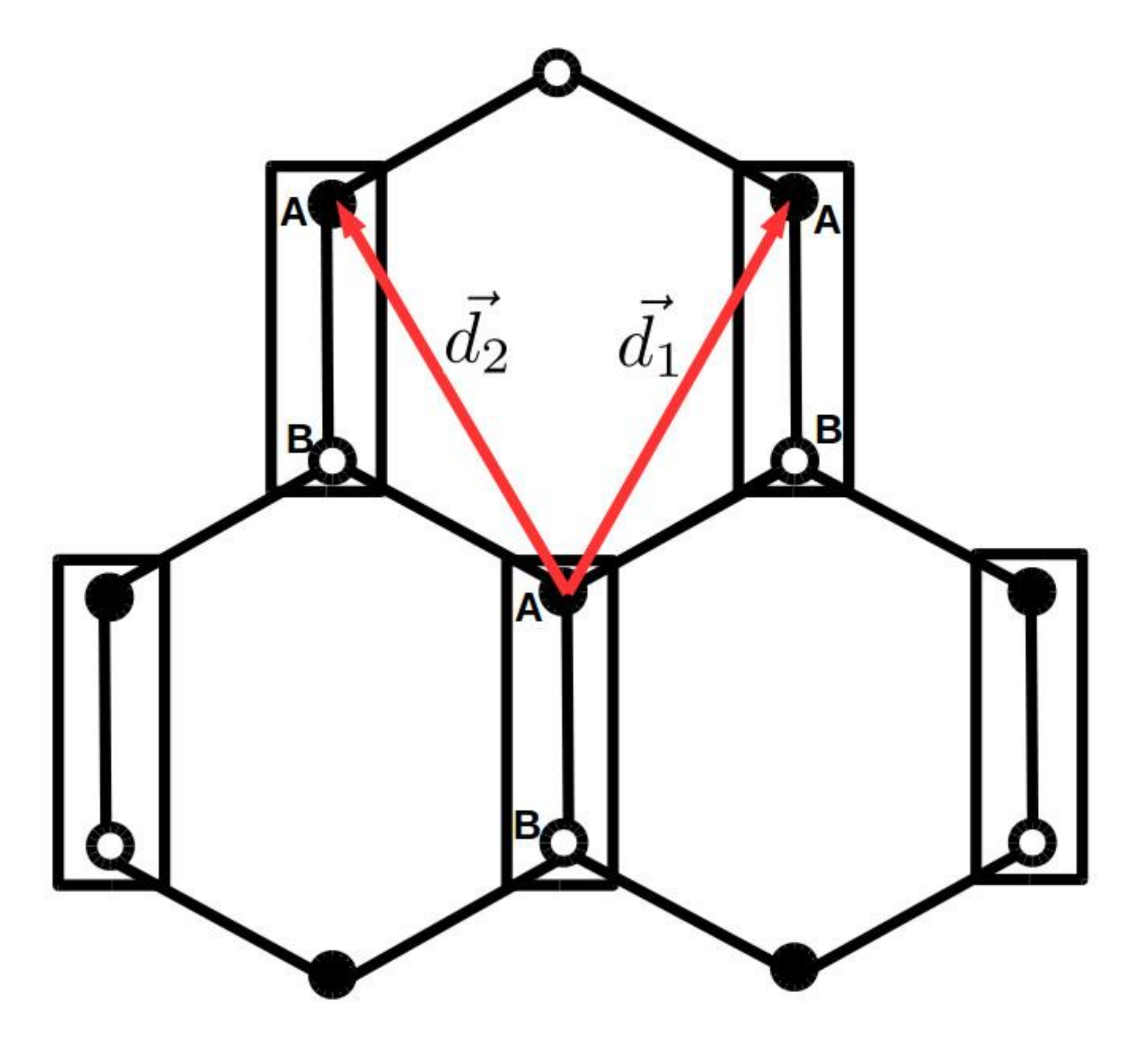}
    \caption{\small The two-point unit cell has been considered along the $z$ bonds. ${\textbf d_1}$ and ${\textbf d_2}$ denotes the two lattice vectors of the Honeycomb lattice.}
\end{figure}
\FloatBarrier

\begin{align}
    c_i=\left\{\begin{array}{l}
    ~f_i+f_i^\dagger~~~~~\forall~ i\in A\\
    i(f_i-f_i^\dagger)~~~\forall~ i\in B\\
    \end{array}\right.
    \label{eq_transformation}
\end{align}

Using this, we get,

\begin{align}
    H_1=&\frac{J_K}{4}\sum_{i}\chi^a\Theta^a_{i,{\bar{p}}}\left[(2f^\dagger_if_i-1)\right.\nonumber\\
    &+(f_if_{i+{\bf d_1}}+f_i^\dagger f_{i+{\bf d_1}}-f_if_{i+{\bf d_1}}^\dagger-f_i^\dagger f_{i+{\bf d_1}}^\dagger)\nonumber\\
    &\left.+(f_if_{i+{\bf d_2}}+f_i^\dagger f_{i+{\bf d_2}}-f_if_{i+{\bf d_2}}^\dagger-f_i^\dagger f_{i+{\bf d_2}}^\dagger)\right]
\end{align}

Using the following Fourier transformation,

\begin{align}
&f_i=\frac{1}{\sqrt{N_b}}\sum_{\bf k} e^{i{\bf k\cdot r}_i}f_{\bf k}\nonumber\\
&\Theta_{i,\bar{p}}^a=\frac{1}{\sqrt{N_b}}\sum_{\bf k} e^{i{\bf k\cdot r}_i}~\Theta_{{\bf k},\bar{p}}^a.
\end{align}

the above Hamiltonian can be written as,

\begin{align}
    H_1=\frac{J_K}{4\sqrt{N_b}}\sum_{{\bf k,k'}}\chi^a\Theta^a_{{\bf k},\bar{p}}&\left[\mathcal{A}_{{\bf k,k'}}f^\dagger_{\bf k'}f_{\bf k'-k}+\mathcal{B}_{{\bf k,k'}}f_{\bf k'}f_{\bf -k-k'}\right.\nonumber\\
    &\left.+\mathcal{C}_{{\bf k,k'}}f_{\bf k'}f^\dagger_{\bf k+k'}+\mathcal{D}_{{\bf k,k'}}f^\dagger_{\bf k'}f^\dagger_{\bf k-k'}\right]
    \label{eq_ffermion}
\end{align}

where

\begin{align}
    \mathcal{A}_{{\bf k,k'}}&=\left[2+e^{i({\bf k'-k})\cdot{\bf d}_1}+e^{i({\bf k'-k})\cdot{\bf d}_2}\right]\nonumber\\
    \mathcal{B}_{{\bf k,k'}}&=\left[e^{-i({\bf k+k'})\cdot{\bf d}_1}+e^{-i({\bf k+k'})\cdot{\bf d}_2}\right]\nonumber\\
    \mathcal{C}_{{\bf k,k'}}&=\left[-e^{-i({\bf k+k'})\cdot{\bf d}_1}-e^{-i({\bf k+k'})\cdot{\bf d}_2}\right]\nonumber\\
    \mathcal{D}_{{\bf k,k'}}&=\left[-e^{i({\bf k'-k})\cdot{\bf d}_1}-e^{i({\bf k'-k})\cdot{\bf d}_2}\right]
\end{align}

The Feynman diagrams for the above interaction is shown in Fig.~\ref{fig:spinphonon}(a) and (b). Now we use the following transformation, which is a standard Bogoliubov rotation, to diagonalize the free Majorana Hamiltonian~\cite{Knolle2}.

\begin{align}
\begin{bmatrix}
    f_{\bf k}\\
    f_{-\bf k}^\dagger
\end{bmatrix}=
\begin{bmatrix}
\cos\theta_{\bf k} & i \sin\theta_{\bf k}\\
i\sin\theta_{\bf k} & \cos\theta_{\bf k}
\end{bmatrix}
\begin{bmatrix}
    a_{\bf k}\\
    a_{-\bf k}^\dagger
\end{bmatrix}
\label{eq_transformation2}
\end{align}

Here, $\tan2\theta_{\bf k}=-\frac{Im S({\bf k})}{Re S({\bf k})}$ and $S({\bf k})=\frac{J_K}{4}(1+e^{i{\bf k\cdot d}_1}+e^{i{\bf k\cdot d}_2})$

Using this transformation, we can rewrite Eq.~\ref{eq_ffermion} in terms of these new fermions ($a$ and $a^\dagger$) which are the normal modes of the zero flux sector. 

\begin{widetext}
\begin{align}
    H_1=\frac{J_K}{4\sqrt{N_b}}\sum_{{\bf k,k'}}\chi^a\Theta^a_{{\bf k},\bar{p}}\Big[\bar{\mathcal{A}}_{{\bf k,k'}}~a_{-{\bf k'}}a_{{\bf k'-k}}+\bar{\mathcal{B}}_{{\bf k,k'}}~a_{-{\bf k'}}a_{{\bf k-k'}}^\dagger
    +\bar{\mathcal{C}}_{{\bf k,k'}}~a^\dagger_{{\bf k'}}a_{{\bf k'-k}}+\bar{\mathcal{D}}_{{\bf k,k'}}~a_{{\bf k'}}^\dagger a_{{\bf k-k'}}^\dagger\Big]
    \label{eq_afermion}
 \end{align}

Here the vertex functions $\bar{\mathcal{A}}_{{\bf k,k}}$, $\bar{\mathcal{B}}_{{\bf k,k'}}$, $\bar{\mathcal{C}}_{{\bf k,k'}}$ and $\bar{\mathcal{D}}_{{\bf k,k'}}$ are given by,

\begin{align}
    &\bar{\mathcal{A}}_{{\bf k,k'}}=i\sin\theta_{-{\bf k'}}\cos\theta_{{\bf k'-k}}\mathcal{A}_{{\bf k,k'}}+\cos\theta_{-{\bf k'}}\cos\theta_{{\bf k'-k}}\mathcal{B}_{{\bf k,-k'}}+i \cos\theta_{-{\bf k'}}\sin\theta_{{\bf k'-k}}\mathcal{C}_{{\bf k,-k'}}-\sin\theta_{-{\bf k'}}\sin\theta_{{\bf k'-k}}\mathcal{D}_{{\bf k,k'}}\nonumber\\
    &\bar{\mathcal{B}}_{{\bf k,k'}}=-\sin\theta_{-{\bf k'}}\sin\theta_{{\bf k'-k}}\mathcal{A}_{{\bf k,k'}}+i\cos\theta_{-{\bf k'}}\sin\theta_{{\bf k'-k}}\mathcal{B}_{{\bf k,-k'}}+\cos\theta_{-{\bf k'}}\cos\theta_{{\bf k'-k}}\mathcal{C}_{{\bf k,-k'}}+i\sin\theta_{-{\bf k'}}\cos\theta_{{\bf k'-k}}\mathcal{D}_{{\bf k,k'}}\nonumber\\
    &\bar{\mathcal{C}}_{{\bf k,k'}}=\cos\theta_{-{\bf k'}}\cos\theta_{{\bf k'-k}}\mathcal{A}_{{\bf k,k'}}+i\sin\theta_{-{\bf k'}}\cos\theta_{{\bf k'-k}}\mathcal{B}_{{\bf k,-k'}}-\sin\theta_{-{\bf k'}}\sin\theta_{{\bf k'-k}}\mathcal{C}_{{\bf k,-k'}}+i\cos\theta_{-{\bf k'}}\sin\theta_{{\bf k'-k}}\mathcal{D}_{{\bf k,k'}}\nonumber\\
    &\bar{\mathcal{D}}_{{\bf k,k'}}=i\cos\theta_{-{\bf k'}}\sin\theta_{{\bf k'-k}}\mathcal{A}_{{\bf k,k'}}-\sin\theta_{-{\bf k'}}\sin\theta_{{\bf k'-k}}\mathcal{B}_{{\bf k,-k'}}+i\sin\theta_{-{\bf k'}}\cos\theta_{{\bf k'-k}}\mathcal{C}_{{\bf k,-k'}}+\cos\theta_{-{\bf k'}}\cos\theta_{{\bf k'-k}}\mathcal{D}_{{\bf k,k'}}
    \label{eq_vertex}
\end{align}
\end{widetext} 

\subsection{The quadratic term \texorpdfstring{$H_2$}{}}

In the zero-flux sector, $H_2$ can be written in terms of Majorana fermions in the following way.

\begin{align}
    H_2=-\frac{J_K}{4}\sum_{i,\alpha}\lambda^{cd}_{\bar{p},\alpha}~\Theta_{i,\bar{p}}^c\Theta_{i,\bar{p}}^d~(ic_ic_{i+\hat{\alpha}})
\end{align}

Now we take a similar approximation as the case of linear coupling considering, $\lambda^{cd}_{\bar{p},\alpha}=\lambda^{cd}$. Further, transforming the Majorana fermions into complex fermions and going to the Fourier basis, we obtain (up to a bare phonon term)

\begin{align}
    H_2=&\frac{J_K}{4N_b}\sum_{{\bf k,k',k''}}\lambda^{cd}~\Theta^c_{{\bf k},\bar{p}}\Theta^d_{{\bf k'},\bar{p}}\times\nonumber\\
    &\left[\mathcal{P}_{{\bf k,k',k''}}f^\dagger_{{\bf k''}}f_{{\bf k''-k-k'}}+\mathcal{Q}_{{\bf k,k',k''}}f_{{\bf k''}}f_{{\bf -k-k'-k''}}\right.\nonumber\\
    &+\left.\mathcal{R}_{{\bf k,k',k''}}f_{{\bf k''}}f^\dagger_{{\bf k+k'+k''}}+\mathcal{S}_{{\bf k,k',k''}}f^\dagger_{{\bf k''}}f^\dagger_{{\bf k+k'-k''}}\right]
\end{align}

where

\begin{align}
    &\mathcal{P_{{\bf k,k',k''}}}=2+e^{i{\bf (k''-k-k')}\cdot {\bf d_1}}+e^{i{\bf (k''-k-k')}\cdot {\bf d_2}}\nonumber\\
    &\mathcal{Q_{{\bf k,k',k''}}}=e^{-i{\bf (k+k'+k'')}\cdot {\bf d_1}}+e^{-i{\bf (k+k'+k'')}\cdot {\bf d_2}}\nonumber\\
    &\mathcal{R_{{\bf k,k',k''}}}=-e^{-i{\bf (k+k'+k'')}\cdot {\bf d_1}}-e^{-i{\bf (k+k'+k'')}\cdot {\bf d_2}}\nonumber\\
    &\mathcal{S_{{\bf k,k',k''}}}=-e^{i{\bf (k''-k-k')}\cdot {\bf d_1}}-e^{i{\bf (k''-k-k')}\cdot {\bf d_2}}
\end{align}

The Feynman diagrams for the above interaction is shown in Fig.~\ref{fig:spinphonon}(c) and (d). Now using Eq.~\ref{eq_transformation2} we get

\begin{widetext}
\begin{align}
    H_2=\frac{J_K}{4N_b}\sum_{{\bf k,k',k''}}\lambda^{cd}~\Theta^c_{{\bf k},\bar{p}}\Theta^d_{{\bf k'},\bar{p}}\left[\bar{\mathcal{P}}_{{\bf k,k',k''}}a_{{\bf -k''}}a_{{\bf k''-k-k'}}+\bar{\mathcal{Q}}_{{\bf k,k',k''}}a_{{\bf -k''}}a^\dagger_{{\bf k+k'-k''}}+\bar{\mathcal{R}}_{{\bf k,k',k''}}a^\dagger_{{\bf k''}}a_{{\bf k''-k-k'}}\right.\nonumber\\
    \left.+\bar{\mathcal{S}}_{{\bf k,k',k''}}a^\dagger_{{\bf k''}}a^\dagger_{{\bf k+k'-k''}}\right]
\end{align}

where

\begin{align}
    &\bar{\mathcal{P}}_{{\bf k,k',k''}}=i\sin\theta_{{\bf -k''}}\cos\theta_{{\bf k''-k-k'}}\mathcal{P}_{{\bf k,k',k''}}+\cos\theta_{{\bf -k''}}\cos\theta_{{\bf k''-k-k'}}\mathcal{Q}_{{\bf k,k',-k''}}\nonumber\\
    &\hspace{6cm}+i\cos\theta_{{\bf -k''}}\sin\theta_{{\bf k''-k-k'}}\mathcal{R}_{{\bf k,k',-k''}}-\sin\theta_{{\bf -k''}}\sin\theta_{{\bf k''-k-k'}}\mathcal{S}_{{\bf k,k',k''}}\nonumber\\
    &\bar{\mathcal{Q}}_{{\bf k,k',k''}}=-\sin\theta_{{\bf -k''}}\sin\theta_{{\bf k''-k-k'}}\mathcal{P}_{{\bf k,k',k''}}+i\cos\theta_{{\bf -k''}}\sin\theta_{{\bf k''-k-k'}}\mathcal{Q}_{{\bf k,k',-k''}}\nonumber\\
    &\hspace{6cm}+\cos\theta_{{\bf -k''}}\cos\theta_{{\bf k''-k-k'}}\mathcal{R}_{{\bf k,k',-k''}}+i\sin\theta_{{\bf -k''}}\cos\theta_{{\bf k''-k-k'}}\mathcal{S}_{{\bf k,k',k''}}\nonumber\\
    &\bar{\mathcal{R}}_{{\bf k,k',k''}}=\cos\theta_{{\bf -k''}}\cos\theta_{{\bf k''-k-k'}}\mathcal{P}_{{\bf k,k',k''}}+i\sin\theta_{{\bf -k''}}\cos\theta_{{\bf k''-k-k'}}\mathcal{Q}_{{\bf k,k',-k''}}\nonumber\\
    &\hspace{6cm}-\sin\theta_{{\bf -k''}}\sin\theta_{{\bf k''-k-k'}}\mathcal{R}_{{\bf k,k',-k''}}+i\cos\theta_{{\bf -k''}}\sin\theta_{{\bf k''-k-k'}}\mathcal{S}_{{\bf k,k',k''}}\nonumber\\
    &\bar{\mathcal{S}}_{{\bf k,k',k''}}=i\cos\theta_{{\bf -k''}}\sin\theta_{{\bf k''-k-k'}}\mathcal{P}_{{\bf k,k',k''}}-\sin\theta_{{\bf -k''}}\sin\theta_{{\bf k''-k-k'}}\mathcal{Q}_{{\bf k,k',-k''}}\nonumber\\
    &\hspace{6cm}+i\sin\theta_{{\bf -k''}}\cos\theta_{{\bf k''-k-k'}}\mathcal{R}_{{\bf k,k',-k''}}+\cos\theta_{{\bf -k''}}\cos\theta_{{\bf k''-k-k'}}\mathcal{S}_{{\bf k,k',k''}}
\end{align}

\end{widetext}

\section{The renormalisation of the Raman active phonons}
\label{A_I}

\subsection{Frequency shift}

In the low temperature QSL regime of the experiment, the spin dynamics is expected to be slower than the optical phonons and hence to the first approximation, the spins can be approximated with their static equal time configuration such that to the leading order, we obtain

\begin{align}
    H_{spin-phonon}\rightarrow \langle H_{spin-phonon}\rangle_S
\end{align}

where $\langle \hat{O}\rangle_S=\frac{Tr(\hat{O}e^{-\beta H_{spin}})}{Tr(e^{-\beta H_{spin}})}$ ($H_{spin}$ is the bare spin Hamiltonian as mentioned in Eq.~\ref{eq_hsp_hph_hspin}) denotes averaging of the equal time spin-correlators over the thermodynamic ensemble. The spin-correlators being time-independent, they now act as a linear and quadratic deformation to $H_{phonon}$. Thus within the Harmonic phonon approximation valid for low temperatures, the linear term $\langle H_1\rangle_S$ does not affect the phonon frequency which is entirely affected by $\langle H_2\rangle_S$ and is given by Eq. \ref{eq_shift} of the main text. In the zero flux approximation, it can be further calculated using free majorana phenomenology.

\begin{align}
    \delta\omega\sim\lambda J_K\sum_{\bf k}\langle\epsilon_{\bf k}(a^\dagger_{\bf k}a_{\bf k}-\frac{1}{2})\rangle_S
    \label{eq_shift_SI}
\end{align}

Clearly, the above expression is directly proportional to the energy of the spin system and therefore always negative. This explains the softening of the phonon with decreasing temperature. Although the zero-flux approximation is not valid to the experimentally relevant temperature regime, the flux excitation only renormalises the above contribution to the frequency.

\begin{figure}[!h]
    \centering
    \includegraphics[width=8.6cm]{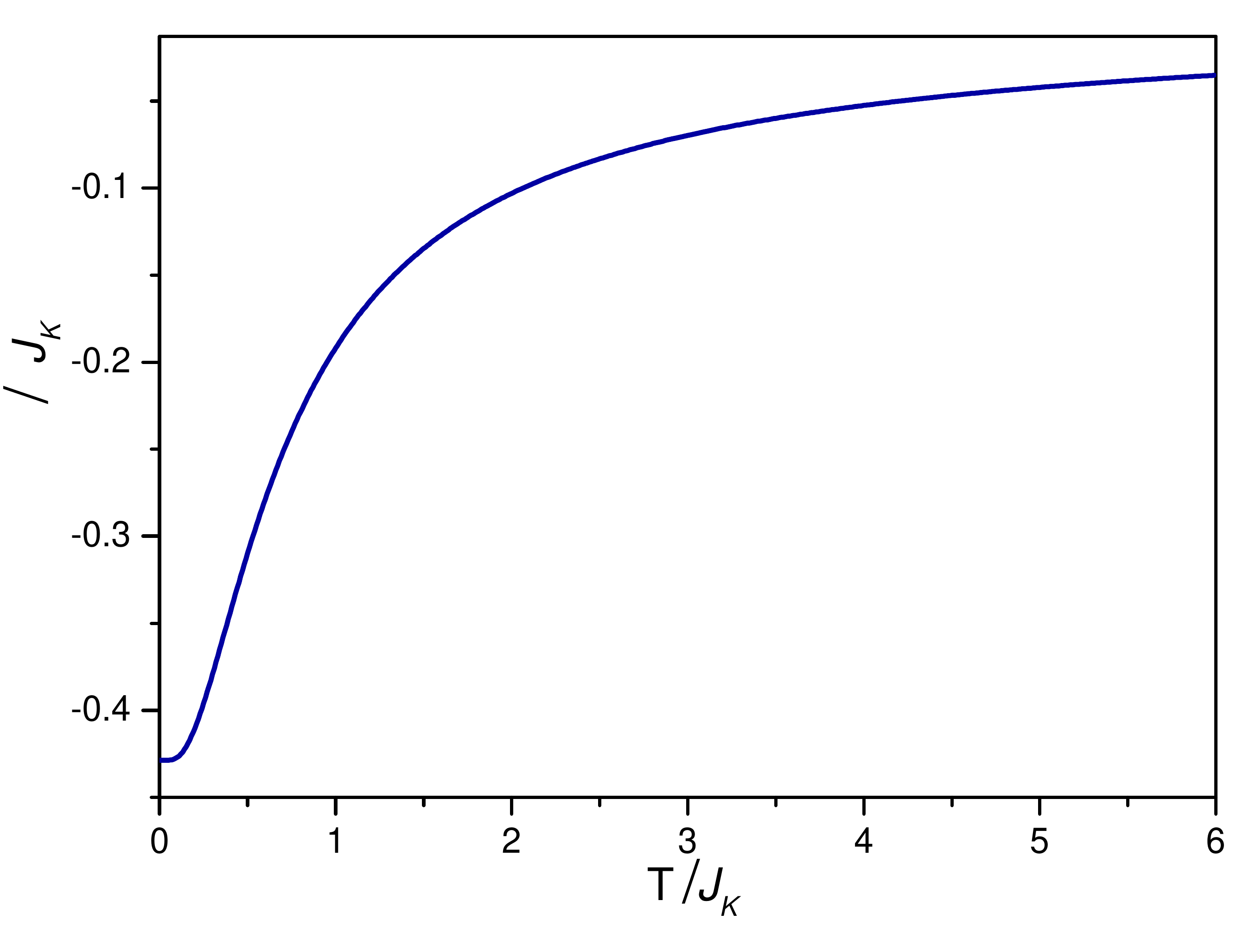}
    \caption{\small Frequency shift is calculated from the average energy of the spin system. At zero temperature, the average energy saturates to $\sim-0.43 J_K$.}
\end{figure}
\FloatBarrier

\subsection{\bf{Linewidth of phonon}}

\begin{figure}[!h]
    \centering
    \includegraphics[width=8.6cm]{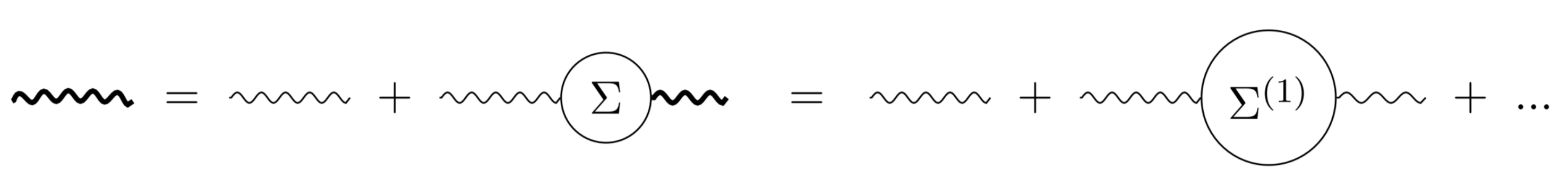}
\caption{\small The thick and thin curly lines denote the dressed and bare phonon Green's function, respectively. In the last diagram, we truncate the series up to one-loop correction.}
    \label{fig_a8}
\end{figure}

Linewidth of the phonon peak is obtained by computing the phonon self-energy ($\Sigma$) defined by the Dyson equation

\begin{align}
	D=D_0+D_0\Sigma D
\end{align}

where, $D_0$ and $D$ are bare and dressed propagator of the phonon, respectively as given in the main text. The above equation is diagrammatically represented in Fig.~\ref{fig_a8}.

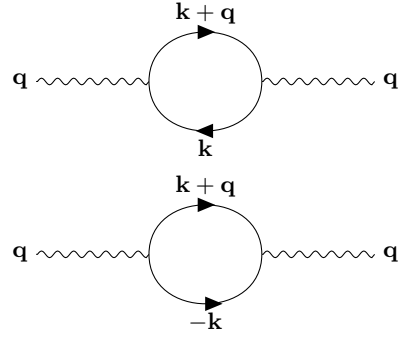
\begin{figure}
    \centering
    {\begin{tikzpicture}
    \begin{feynman}
    \vertex  (a);
    \vertex [left=of a] (b);
    \vertex [right=of a, label=right:\({\bf q}\)] (f1);
    \vertex [left=of b, label=left:\({\bf q}\)] (f3); \diagram* {
    (f3) -- [boson] (b) -- [fermion, half left,edge label=\({\bf k+q}\)] (a) -- [boson] (f1), (a)--[fermion, half left, edge label=\({\bf k}\)](b)
        };
    \end{feynman}
    \end{tikzpicture}}
    {\begin{tikzpicture}
    \begin{feynman}
    \vertex  (a);
    \vertex [left=of a] (b);
    \vertex [right=of a, label=right:\({\bf q}\)] (f1);
    \vertex [left=of b, label=left:\({\bf q}\)] (f3); \diagram* {
    (f3) -- [boson] (b) -- [fermion, half left,edge label=\({\bf k+q}\)] (a) -- [boson] (f1), (a)--[anti fermion, half left,edge label=\({\bf -k}\)](b)
        };
    \end{feynman}
    \end{tikzpicture}}    
    \caption{The solid lines represent the propagator for 'a'-fermions.}
    \label{fig_a9}
\end{figure}

This series can be computed perturbatively under the assumption that spin-phonon coupling is the weakest energy scale of the problem. Here we go up to one loop contribution. At this order, the self-energy can be computed from the Feynman diagrams shown in Fig.~\ref{fig_a9}.

\begin{widetext}
\begin{align}
	\Sigma({\bf q},i\omega)&\sim-\frac{\chi^2J_K^2}{N_b\beta}\sum_{\omega_m,{\bf k}}\left[M_{{\bf k+q,k}}G({\bf k+q},-i\omega-i\omega_m)G(-{\bf k},i\omega_m)+
	M_{{\bf-k-q,-k}}G({\bf k+q},i\omega-i\omega_m)G(-{\bf k},i\omega_m)\right.\nonumber\\
	&~~~~~~~~~~~~~~~~~~~~~~~+\left.N_{{\bf k+q,k}}G({\bf k+q},i\omega+i\omega_m)G({\bf k},i\omega_m)+N_{{\bf-k-q,-k}}G({\bf k+q},-i\omega+i\omega_m)G({\bf k},i\omega_m)\right]\nonumber\\
	&=\frac{\chi^2J_K^2}{N_b\beta}\sum_{\bf k}\sum_{\omega_m}\left[M_{\bf k+q,k}\frac{1}{i\omega+i\omega_m+\epsilon_{\bf k+q}}\frac{1}{i\omega_m-\epsilon_{\bf k}}-M_{\bf -k-q,-k}\frac{1}{i\omega-i\omega_m-\epsilon_{\bf k+q}}\frac{1}{i\omega_m-\epsilon_{\bf k}}\right.\nonumber\\
	&~~~~~~~~~~~~~~~~~~~~~~~\left.-N_{\bf k+q,k}\frac{1}{i\omega+i\omega_m-\epsilon_{\bf k+q}}\frac{1}{i\omega_m-\epsilon_{\bf k}}+N_{\bf -k-q,-k}\frac{1}{i\omega-i\omega_m+\epsilon_{\bf k+q}}\frac{1}{i\omega_m-\epsilon_{\bf k}}\right]
\end{align}

where,

\begin{align*}
&G({\bf k},i\omega)=-\int_0^\beta d\tau\langle\mathcal{T}a_{\bf k}(\tau)a_{\bf k}^\dagger(0)\rangle_0 e^{i\omega\tau}=\frac{1}{i\omega-\epsilon_{\bf k}}~~~~~~~~~~~~~\\
&M_{\bf k+q,k}=\bar{\mathcal{A}}_{\bf -q,k}\bar{\mathcal{D}}_{\bf q,-k}+\bar{\mathcal{A}}_{\bf -q,-k-q}\bar{\mathcal{D}}_{\bf q,-k}+\bar{\mathcal{D}}_{\bf q,-k}\bar{\mathcal{A}}_{\bf -q,k}+\bar{\mathcal{D}}_{\bf q,k+q}\bar{\mathcal{A}}_{\bf -q,k}\\
&N_{\bf k+q,k}=\bar{\mathcal{B}}_{\bf -q,-q-k}\bar{\mathcal{B}}_{\bf q,-k}+\bar{\mathcal{C}}_{\bf -q,k}\bar{\mathcal{B}}_{\bf q,-k}+\bar{\mathcal{B}}_{\bf -q,k}\bar{\mathcal{C}}_{\bf q,-k}+\bar{\mathcal{C}}_{\bf q,k+q}\bar{\mathcal{C}}_{\bf -q,k}
\end{align*}

\end{widetext}

To perform the Matsubara frequency summation, we define

\begin{align}
I_1=\lim_{R\rightarrow\infty}\frac{1}{2\pi i}\oint \frac{1}{e^{\beta z}+1}\frac{1}{z+i\omega+\epsilon_{\bf k+q}}\frac{1}{z-\epsilon_{\bf k}}\\
I_2=\lim_{R\rightarrow\infty}\frac{1}{2\pi i}\oint \frac{1}{e^{\beta z}+1}\frac{1}{z+i\omega-\epsilon_{\bf k+q}}\frac{1}{z-\epsilon_{\bf k}}
\end{align}

Here, the contour is chosen to be the circle with radius $R$ and the radius is sent to $\infty$.

%\squeezetable
\begin{table}[!ht]
\begin{center}
\begin{tabular}{|m{1.5cm}|m{2.4cm}||m{1.5cm}|m{2.4cm}|@{}m{0pt}@{}}
\hline
\multicolumn{2}{|c||}{$I_1$} & \multicolumn{2}{|c|}{$I_2$} &\\
\hline\hline
~~~ Poles & ~~~~~~~ Residue & ~~~ Poles & ~~~~~~~ Residue &\\ [8pt]
\hline
 $~~~~i\omega_m$ & $-\frac{1}{\beta}\sum_{\omega_m}\frac{1}{i\omega_m-\epsilon_{\bf k}}$ & $~~~~i\omega_m$ & $-\frac{1}{\beta}\sum_{\omega_m}\frac{1}{i\omega_{m}+\epsilon_{\bf k}}$ &\\[8pt]  
 $ $ & $~~~\times\frac{1}{i\omega+i\omega_m+\epsilon_{\bf k+q}}$ & $ $ & $~~~\times\frac{1}{i\omega+i\omega_m-\epsilon_{\bf k+q}}$ &\\[8pt]
\hline
 $~~~~\epsilon_{\bf k}$ & $~~~~~\frac{n_F(\epsilon_{\bf k})}{i\omega+\epsilon_{\bf k}+\epsilon_{\bf k+q}}$ &  $~~~~\epsilon_{\bf k}$ & $~~~~~\frac{n_F(\epsilon_{\bf k})}{i\omega+\epsilon_{\bf k}-\epsilon_{\bf k+q}}$ &\\[8pt]  
\hline     
 $-i\omega-\epsilon_{\bf k+q}$ &  $~~-\frac{1-n_F(\epsilon_{\bf k+q})}{i\omega+\epsilon_{\bf k+q}+\epsilon_{\bf k}}$   &   $-i\omega+\epsilon_{\bf k+q}$ &  $~~-\frac{n_F(\epsilon_{\bf k+q})}{i\omega+\epsilon_{\bf k}-\epsilon_{\bf k+q}}$ & \\[8pt]  
\hline 
\end{tabular}
\end{center}
\end{table}

Now, both $I_1$ and $I_2$ vanishes as $R\rightarrow\infty$. Therefore,

\begin{align*}
	-\frac{1}{\beta}\sum_{\omega_m}\frac{1}{i\omega+i\omega_m+\epsilon_{\bf k+q}}\frac{1}{i\omega_m-\epsilon_{\bf k}}=\frac{1-n_F(\epsilon_{\bf k+q})-n_F(\epsilon_{\bf k})}{i\omega+\epsilon_{\bf k}+\epsilon_{\bf k+q}}\\
	\frac{1}{\beta}\sum_{\omega_m}\frac{1}{i\omega+i\omega_m-\epsilon_{\bf k+q}}\frac{1}{i\omega_m+\epsilon_{\bf k}}=\frac{n_F(\epsilon_{\bf k})-n_F(\epsilon_{\bf k+q})}{i\omega+\epsilon_{\bf k}-\epsilon_{\bf k+q}} ~~~~~
\end{align*}

Hence,

\begin{widetext}

\begin{align}
\Sigma(q,i\omega)\sim-\frac{\chi^2J_K^2}{N_b}\sum_{\bf k}\left[\left(1-n_F(\epsilon_{\bf k})-n_F(\epsilon_{\bf k+q})\right)\left(\frac{M_{\bf k+q,k}}{i\omega+\epsilon_k+\epsilon_{\bf k+q}}-\frac{M_{\bf -k-q,-k}}{i\omega-\epsilon_{\bf k}-\epsilon_{\bf k+q}}\right)\right.\nonumber\\
+\left.\left(n_F(\epsilon_{\bf k})-n_F(\epsilon_{\bf k+q})\right)\left(\frac{N_{\bf k+q,k}}{i\omega+\epsilon_{\bf k}-\epsilon_{\bf k+q}}-\frac{N_{\bf -k-q,-k}}{i\omega+\epsilon_{\bf k+q}-\epsilon_{\bf k}}\right)\right]
\label{eq_self-energy}
\end{align}

\end{widetext}

In this work, we consider the ${\bf q}\rightarrow 0$ limit which is relevant to the Raman scattering. At this limit, the second part of the above expression with the factor ($n_F(\epsilon_{\bf k})-n_F(\epsilon_{\bf k+q})$) vanishes. Finally, we Taylor expand $M_{\bf k+q,k}$ in powers of momentum and truncate the series upto the first non-zero term which is momentum independent. This gives the leading temperature dependence of the self-energy. We take the imaginary part of the above expression after doing the analytic continuation. Using the identity $Im\left[\frac{1}{x-x_0+i0^+}\right]=-\pi \delta(x-x_0)$, we obtain

\begin{align}
    Im[\Sigma({\bf q},\omega+i0^+)]\sim\frac{\pi\chi^2J_K^2}{N_b}\sum_{\bf k}(1-n_F(\epsilon_{\bf k})-n_F(\epsilon_{\bf k+q}))\nonumber\\
    \times\Big[\delta(\omega+\epsilon_{\bf k}+\epsilon_{\bf k+q})-\delta(\omega-\epsilon_{\bf k}-\epsilon_{\bf k+q})\Big]
    \label{eq_imself}
\end{align}
We further perform the momentum integral in Eq.~\ref{eq_imself} numerically for ${\bf q}=0$ considering the free fermionic dispersion to be $\epsilon_{\bf k}=2\mid S({\bf k})\mid$~\cite{Knolle2}. 

\begin{figure}[!h]
    \centering
    \includegraphics[width=8.6cm]{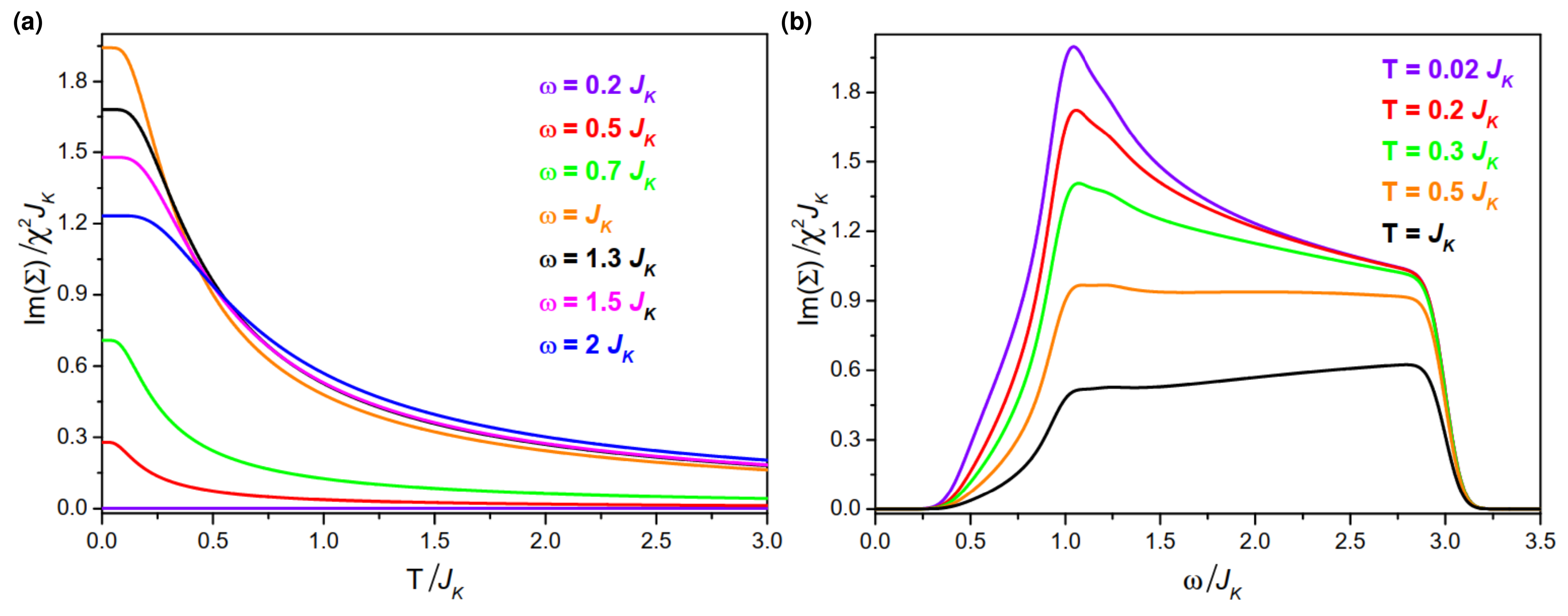}
    \caption{\small Variation of phonon linewidth with ({\bf a}) temperature and ({\bf b}) frequency scaled w.r.t. the Kitaev coupling \textit{J$_K$}.}
    \label{fig_linewidth}
\end{figure}
\FloatBarrier

The peak in Fig.~\ref{fig_linewidth}(b) actually corresponds to the peak in density of states of free Majoranas. In the experimental temperature regime, the flux excitations further renormalise the density of states. But it does not change the qualitative features of the above plots. Note that real part of Eq.~\ref{eq_self-energy} also contributes to the renormalisation of the frequency shift. But we neglect this second order contribution compared to Eq.~\ref{eq_shift_SI}.\vspace{0.8cm}

\subsection{Fitting with free Majorana calculation for the M1 mode}

\begin{figure}[!h]
    \centering
    \includegraphics[width=8.6cm]{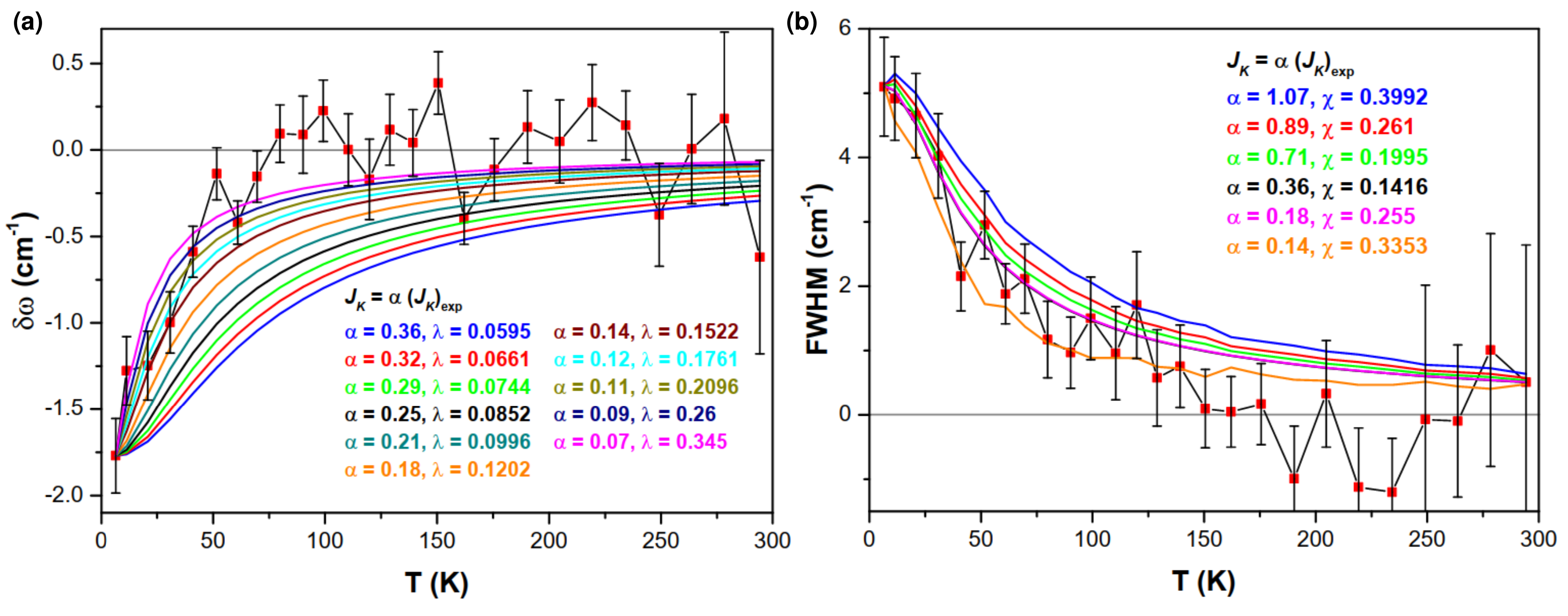}
    \caption{\small Temperature dependence of {\bf (a)} frequency shift and {\bf (b)} FWHM for the M1 mode. Red squares are the experimental data after subtracting the anharmonic contribution. The smooth lines represent the theoretical curves at different values of coupling constant. In {\bf (b)}, we plot only till $J_K$ = 40 K considering the finite bandwidth ($\sim 3J_K$) effect as shown in Fig.~\ref{fig_linewidth} {\bf (b)}.} 
    \label{fig_fitting}
\end{figure}

For completion we compare our free Majorana results with the experiments for the $M1$ mode as shown in Fig.~\ref{fig_fitting}.

\end{document}